\documentclass[twocolumn,showkeys,preprintnumbers,amsmath,amssymb]{revtex4}

\usepackage{graphicx}
\usepackage{dcolumn}
\usepackage{bm}

\begin{document}

\title{On-chip thermometry for microwave optomechanics implemented in a nuclear demagnetization cryostat}
\author{ X. Zhou$^{*,**}$, D. Cattiaux$^{*}$, R. R. Gazizulin$^{*}$, A. Luck$^{*}$, O. Maillet$^{*}$, T. Crozes$^{*}$, J-F. Motte$^{*}$, O. Bourgeois$^{*}$, A. Fefferman$^{*}$ and E. Collin$^{*,\dag}$}

\address{(*) Univ. Grenoble Alpes, Institut N\'eel - CNRS UPR2940, 
25 rue des Martyrs, BP 166, 38042 Grenoble Cedex 9, France \\
          (**) Since 01/10/2017: IEMN, Univ. Lille - CNRS UMR8520, 
Av. Henri Poincar\'e, Villeneuve d'Ascq 59650, France 
}


\date{\today}

\begin{abstract}

We report on 
microwave optomechanics measurements performed on a nuclear adiabatic demagnetization cryostat, whose 
temperature is determined by accurate 
thermometry from below 500$~\mu$K to about 1$~$Kelvin. 
We describe a method for accessing the on-chip temperature, building on the blue-detuned parametric instability and a standard microwave setup. 
The capabilities and sensitivity of both the experimental arrangement and the developed technique are demonstrated with a very weakly coupled silicon-nitride doubly-clamped beam mode of about 4$~$MHz and a niobium on-chip cavity resonating around 6$~$GHz.
We report on an unstable intrinsic driving force in the coupled microwave-mechanical system acting on the mechanics that appears below typically 100$~$mK. The origin of this phenomenon remains unknown, and deserves theoretical input.
It prevents us from performing reliable experiments below typically 10-30$~$mK; however {\it no evidence} of thermal decoupling is observed, and we propose that the same features should be present in {\it all devices} sharing the microwave technology, at different levels of strengths. 
We further demonstrate empirically how most of the unstable feature can be annihilated, and speculate how the mechanism could be linked to atomic-scale two level systems. 
The described microwave/microkelvin facility is part of the EMP platform
, and shall be used for further experiments within and {\it below} the millikelvin range. 

\end{abstract}

\keywords{Mechanics, Condensed Matter Physics, Quantum Physics}

\maketitle

\section{Introduction}

Advances in clean-room technologies within the last decades make it possible to create mechanical elements with one (or more) dimensions below a micron, namely NEMS (Nano-Electro-Mechanical Systems) \cite{roukescleland}. These objects can be embedded in electronic circuits, and today even within {\it quantum} electronic circuits \cite{cleland2010,quantelecsimmonds,quantelec2}. Indeed a breakthrough has been made with microwave optomechanics, essentially shifting the concepts of optomechanics into the microwave domain \cite{lehnert2008}. These quantum-{\it mechanical} devices can then be thought of as a new resource for quantum electronics and quantum information processing, with the realization of e.g. quantum-limited 
 optical-photon/microwave-photon converters and non-reciprocal microwave circuits 
\cite{convOskar,cindy,clerknonrec,simmondsamplifnonrecip,kippenbergamplifnonrecip,jfink}.
Profound quantum concepts are also under study, with for instance mechanical motion squeezing \cite{schwavsciencesqueeze} and 
recently mechanical entanglement of {\it separate} objects \cite{sillanpaaintrique}.

In order to operate at the quantum limit, %
the mechanical mode in use has to be initially in its quantum ground state; for experiments on microwave optomechanics based on megahertz motion, this shall rely on a strong red-detuned pump signal that {\it actively} cools the mode, with an environment remaining ``hot'' 
\cite{kippenbergamplifnonrecip,sillanpaamultimode,schwabtones}. 
For most applied issues this is not a problem, even if it brings an additional complexity with a microwave tone that has to be kept on, which could lead to heating of the circuit and mixing with other signals involved.
However, for all experiments aiming at a careful characterization of the mechanical decoherence due to the thermal bath, this is impractical: the whole system is required to be in thermodynamic equilibrium.
As such, theoretical proposals testing the grounds of quantum mechanics have been released in the literature, both on microwave setups using quantum circuits \cite{ArmourBlencowe} and conventional optomechanics (i.e. with lasers) \cite{bouwmeesterNJP}. 

Indeed, the first attempt to use nuclear adiabatic demagnetization for optomechanics is due to Dirk Bouwmeester, building on the technology of Leiden Cryogenics \cite{KlecknerThesis}. 
This ambitious setup was aimed at cooling a {\it macroscopic} moving mirror actuated/detected by optical means. 
On the other hand, a microwave-based optomechanical system hosting a NEMS is much less demanding; the heat loads from both the photon field and the internal heat release of materials are much less. As such, these setups have already proven to be compatible with dilution (millikelvin) technology \cite{cleland2010,quantelecsimmonds,quantelec2,lehnert2008,kippenbergamplifnonrecip,schwavsciencesqueeze,sillanpaaintrique}.

\begin{figure*}[t!]		 
\center
			 \includegraphics[scale=0.52]{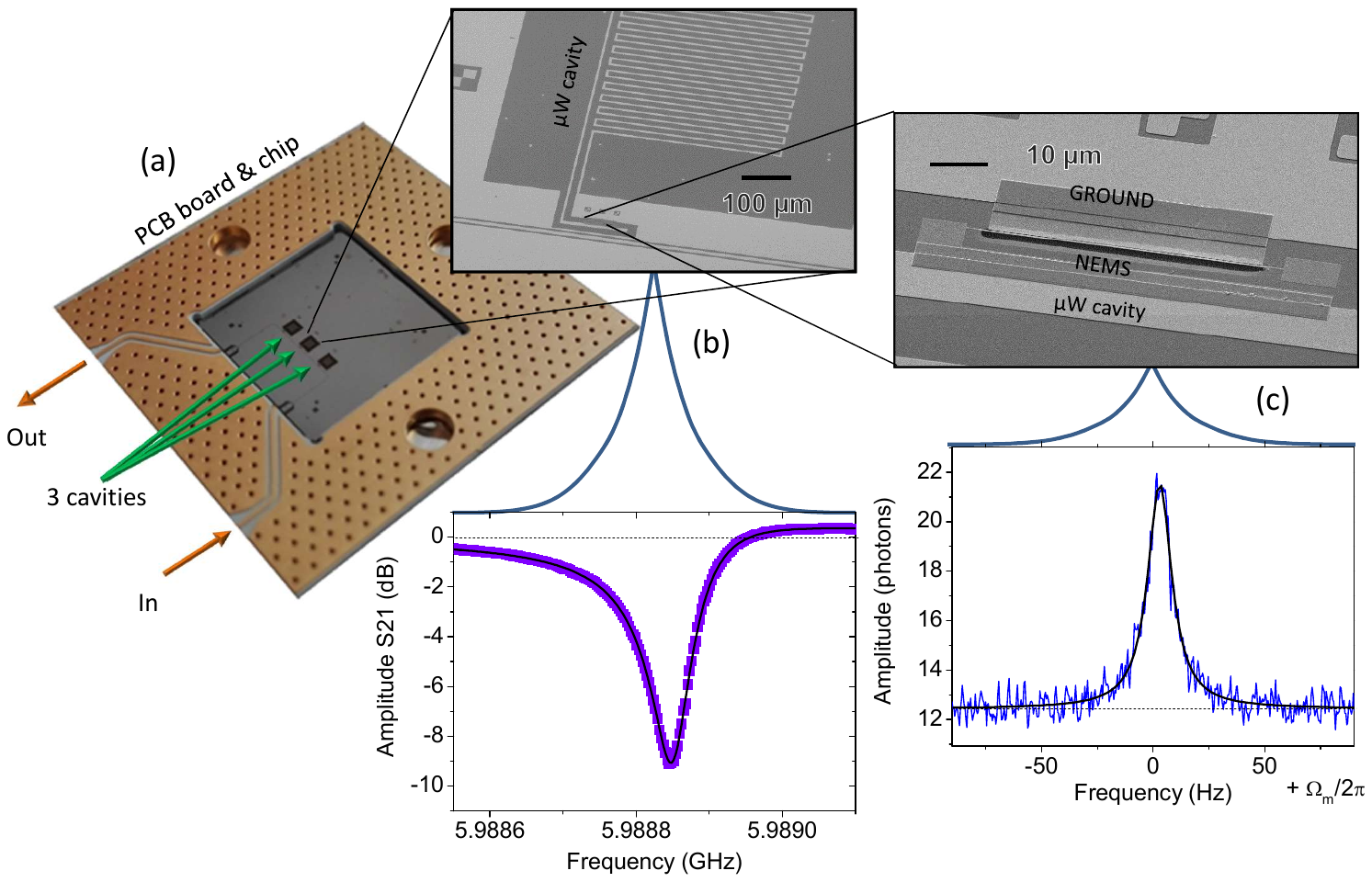}
			\vspace*{-0.9cm}
			\caption{
			Chip arrangement. (a): chip before bonding in PCB (microwave circuit board), displaying a coplanar transmission line coupled to 3 microwave cavities. (b): cavity and resonance [data violet squares, black line is a fit to Eq. (\ref{eq_S21})] at 210$~$mK with blue-detuned pump power 
			7$~$nW and probe power 
			25$~$fW. (c): NEMS beam structure and resonance spectrum measured on the Stokes peak for blue-detuned pump power 
			0.5$~$nW at same temperature (negligible optomechanical back-action, data blue line and black line is a Lorentzian fit with $\Omega_m\approx 2 \pi \times 3.79~$MHz, see Appendix \ref{optomechs} for details).}
			\label{fig_1}
		\end{figure*}

We have thus built a microwave platfrom for optomechanics on a nuclear adiabatic demagnetization cryostat. The microwave circuitry has been kept as basic as possible so far for demonstration purposes; no JPA (Josephson Parametric Amplifier) has been used, and we rely only on {\it intrinsic} properties of optomechanics for the measurement \cite{AKMreview}. The cryostat reaches temperatures below 500$~\mu$K, and is equipped with 
accurate
thermometry from the lowest temperatures up to about 1$~$K: using a noise SQUID-based (Superconducting QUantum Interference Device) thermometer \cite{noiseJohn} plus a $^3$He-fork thermometer \cite{Rob3Hefork}.

We report on 
measurements realized on this platform with a very weakly coupled doubly-clamped beam flexural mode embedded in an on-chip cavity.
Beam-based devices are indeed a tool of choice for applications like ultimate sensing (e.g. single molecule mass spectrometry \cite{roukesmass}), 
while the smallness of the photon-phonon coupling enables demonstrating the sensitivity of the methods used: we build on the optomechanical blue-detuned pumping instability to extract the {\it on-chip} temperature. 
On the other hand, popular drumhead devices achieve routinely much higher couplings \cite{schwavsciencesqueeze,quantelecsimmonds} (from $\times 10$ to $\times 100$); the very same methods can then be applied with much less power, therefore limiting heating problems.
We report not only on the thermalization of the mechanical mode itself, but also on the bulk of the mechanical element with the temperature of its constitutive TLSs (Two-Level Systems). 

Below about 100$~$mK, the mechanical device starts to be self-driven out-of-equilibrium by a strong stochastic force of unknown origin. We report on this phenomenon 
with a detailed account of the observed features, which have been confirmed in different laboratories but never documented to our knowledge \cite{sillanpaPrivCom,schwabPHD,contactLehnert}. 
This thorough description is calling for theoretical understanding, and we speculate on the possible ingredients that could be underlying this effect.
We demonstrate that the strongest events can be canceled by applying a DC voltage onto the transmission line.
 However, for beam-based devices the remaining (small) events seem to limit the measurement capabilities to about 10$~$mK, roughly the lowest achievable temperature for dilution cryostats. 
This is actually an order of magnitude better than what has been reported so far for doubly-clamped beams \cite{sillanpaamultimode,schwabrocheleau,TeufelBEAM}. 
Looking also at the spread in the reported thermalization temperatures of drum-like devices, obtained in {\it completely similar} conditions, we suspect that this genuine phenomenon should be present in all devices sharing the same microwave readout scheme and constitutive materials, 
at different levels of expression.
In this respect, the present description is extremely relevant for the community, and calls for further experiments on other types of devices {\it below} 10$~$mK.

By carefully characterizing the self-heating due to the {\it continuously} injected microwave power, we infer that for the extremely weakly coupled mode used here the technique is suitable down to temperatures of the order of $1-2~$milliKelvin; for much larger couplings (like in drum structures) where less power is needed, in the {\it absence} of any uncontrolled intrinsic drive the technique should be functional down to the lowest achievable temperatures.
As a comparison, similar experiments in the millikelvin range using optics could only be performed in pulsed mode \cite{DavisTLS}.

\section{Experiment}
\label{experiment}

The microwave setup that we have chosen for this demonstration is relatively basic, and resembles the one described in Ref. \cite{lehnert2008}. 
Details on the microwave wiring can be found in Appendix \ref{setups}.
The measurement is performed in transmission, with a coplanar transmission line coupled on-chip to microfabricated microwave cavities (see Fig. \ref{fig_1} left). The practical aspect of this design is multiplexing: in Fig. \ref{fig_1} one cavity hosts our on-chip thermometer while the others can be used for more complex experiments.

The microwave cavities are realized through laser lithography and RIE (Reactive Ion Etching) of a 120$~$nm thick layer of niobium (Nb). 
A typical resonance curve at about $\omega_c/(2 \pi) \approx 6~$GHz is shown in Fig. \ref{fig_1} center, displaying a (bi-directional) coupling rate of $\kappa_{ext} /(2 \pi)\approx 100~$kHz and a total damping rate of about $\kappa_{tot}/(2 \pi) \approx 150~$kHz.
The NEMS mechanical element we use as on-chip thermometer is made from 80$~$nm thick high-stress silicon-nitride (SiN, 0.9$~$GPa), grown on top of silicon. It is a 50$~\mu$m long doubly-clamped beam of width 300$~$nm. It is covered by a 30$~$nm layer of aluminum (Al), capacitively coupled to the cavity through a 100$~$nm gap.
The aluminum part has been patterned using standard e-beam lithography and lift-off, while the beam was released through RIE etching of the silicon-nitride followed by a selective XeF$_2$ silicon etching \cite{KunalsJLTP}.
The silicon-nitride has not been removed below the niobium layer.
The mechanical resonance of the first flexural mode we use is shown in Fig. \ref{fig_1} right, around $\Omega_m/(2 \pi) \approx 4~$MHz, with damping rate $\gamma_m/(2 \pi)$ of order 10$~$Hz. 
Such a mechanical mode still hosts about $n_{thermal} \approx 6~$phonons around 1$~$mK, large enough to be considered in the {\it classical} limit and thus well adapted to mode thermometry.

The microwave setup has been carefully calibrated by electronic means; we therefore quote injected powers in Watts applied on-chip, and detected spectra in Watts/Hz (expressed in  {\it number of 6$~$GHz photons}) at the output port of the chip. 
These units are adapted to both issues of heat leaks (cryogenics) and signal-to-noise (electonics) discussed in this Article. 
Two setups have been used: a dry commercial BlueFors$^{\textregistered}\,$ (BF) machine for preliminary experiments (base temperature 7.5$~$mK), and then 
the home-made Grenoble nuclear adiabatic demagnetization cryostat (operated down to 400$~\mu$K) \cite{paperYuriSatge}. 
Particular care has been taken in the construction of the two cryostats' temperature scales, and details are given in Appendix \ref{setups}.

		\begin{figure}
		\centering
	\includegraphics[width=8.5cm]{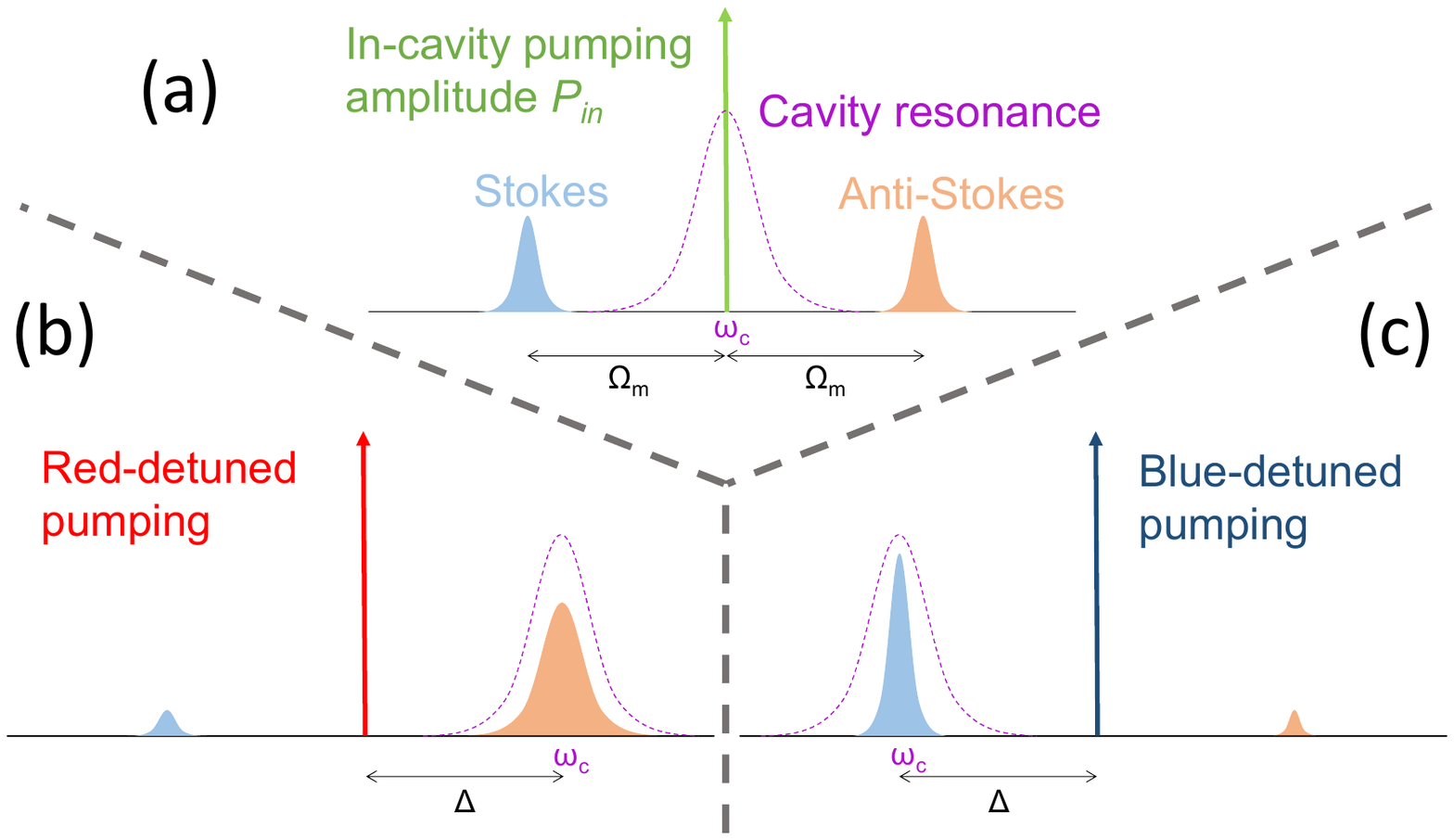}
	\vspace*{-0.5cm}
			\caption{
			Optomechanical schemes used (drawing not to scale, frequency domain). (a): ``in-cavity'' pumping, a strong tone (in green) is applied at the cavity resonance and the two (equivalent) sidebands are measured. (b) Red-detuned pumping: a strong microwave signal is applied  (in red) at a frequency detuned from the cavity $\omega_c$ by  $\Delta = -\Omega_m$, and we measure the so-called anti-Stokes peak at $\omega_c$. (c) Blue-detuned pumping: a strong pump tone  (in blue) is applied at frequency $\omega_c+\Omega_m$, while we measure the Stokes peak.}
			\label{fig_2}
		\end{figure}
			
The opto-mechanical interaction arises from the force exerted by light onto movable objects \cite{AKMreview}. 
It corresponds to the transfer of momentum carried by light (i.e. photon particles) to the surfaces on which it reflects. 
In a cavity design with one movable mirror, the retarded nature of this so-called radiation pressure force (due to the finite lifetime of light inside the cavity) leads to damping or anti-damping of the motion, depending on the frequency detuning of the input light with respect to the cavity \cite{AKMreview}. 

Our moving end mirror is the NEMS beam capacitively coupled to the microwave resonator of Fig. \ref{fig_1} (the cavity) \cite{lehnert2008}.
The standard schemes we use are illustrated in Fig. \ref{fig_2}, with an input wave of power $P_{in}$ at frequency $\omega_c+\Delta$. We are in the so-called resolved-sideband regime, with $\gamma_m, \kappa_{tot} \ll \Omega_m$ \cite{AKMreview}. 
When we pump power in the system exactly at the frequency of the cavity $\omega_c$ (detuning $\Delta=0$), the mechanical motion leads 
to a phase shift of the light \cite{AKMreview}. The Brownian motion of the mechanical element thus imprints two equivalent sidebands in the spectrum that we can measure.
For a red-detuned pump (frequency detuning $\Delta = -\Omega_m$, Fig. \ref{fig_2}), energy quanta from the mechanics (phonons) can be transferred to the optical field (photons). This leads to the well-known sideband cooling technique \cite{AKMreview,firstopticsSBcool,microwaveSBCoolLehnertPRL}. The so-called anti-Stokes sideband peak in the spectrum is favored, and the mechanical mode is {\it damped} by light.
On the other hand for a blue-detuned pump ($\Delta = +\Omega_m$, Fig. \ref{fig_2}), the optical field generates a parametric instability in the mechanics \cite{AKMreview,clerkmarquardtharris2010}. The Stokes sideband peak is favored, 
and the mechanics is amplified through {\it anti-damping}. Eventually one can reach at high pumping powers the self-oscillation regime \cite{kippenbergOscillVahalaPRL2005,delftsteeleOscill}. \\

		\begin{figure}[h!]
		\centering
	\includegraphics[width=11cm]{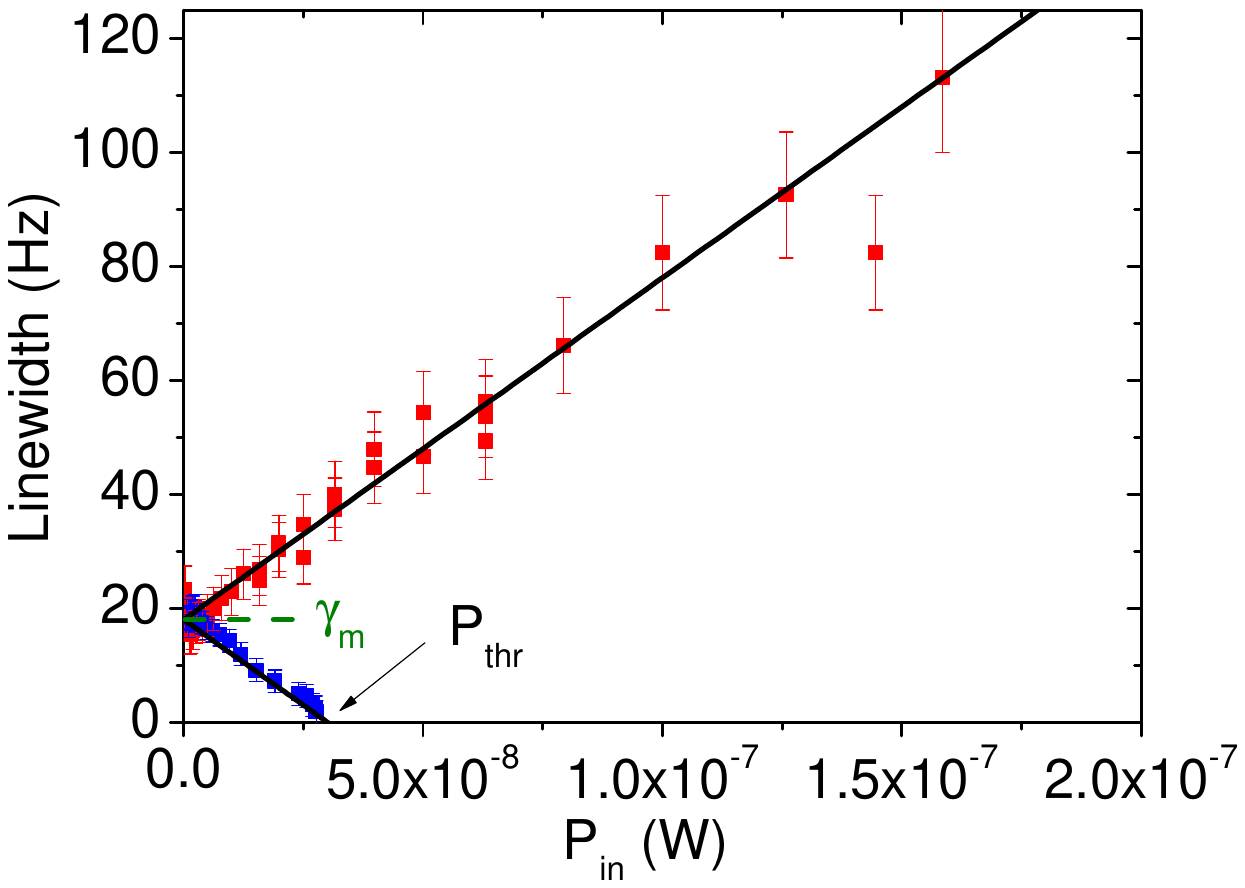}
	\vspace*{-2cm}
			\caption{
			Effective damping/anti-damping $\gamma_{ef\!f}$ measured for the blue and the red-detuned pumping schemes as a function of power (at 210$~$mK, blue and red squares respectively). The slope of the fit (black line) leads to the definition of $g_0$, while the $P_{in} \approx 0$ corresponds to $\gamma_m$. The arrow indicates the position of the threshold $P_{thr}$ towards self-sustained oscillations, which is simply proportional to $\gamma_m$ (see text).}
			\label{fig_3}
		\end{figure}

We start by calibrating the optomechanical interaction.	
The optical damping and anti-damping are linear in applied power $P_{in}$, see Fig. \ref{fig_3}. From a fit [Eq. (\ref{eq1}) below], we can infer the so-called single photon coupling strength $g_0 = \frac{1}{2} \omega_c \frac{1}{C} \frac{d C}{d x} x_{zpf}$ with $x_{zpf}$ the zero-point-motion \cite{AKMreview}.
 This is essentially a {\it geometrical} parameter, arising from the modulation $\frac{d C}{d x}$ of the microwave mode capacitance $C$ by the beam motion \cite{lehnert2008}.
We find $g_0/(2 \pi) \approx 0.55 \pm 0.1~$Hz, corresponding to the out-of-plane flexure.
This coupling is particularly small, the idea being to take advantage of that to demonstrate the sensitivity of our method.
The magnitude of the output power is fit to theory \cite{AKMreview}, leading to a calibration of the measured phonon mode population/temperature [performed at 210$~$mK, parameter $\cal M$ in Eq. (\ref{eq2}) and Fig. \ref{fig_fit}]. 
More details on the optomechanics measurements can be found in Appendix \ref{optomechs}. 

\section{Method}

The method we propose builds on the parametric instability of the blue-detuned pumping scheme.
When the pump tone is applied at $\omega_c$, the size of the two equivalent sideband peaks (their measured area ${\cal A}_0$, in photons/s) is simply proportional to injected power $P_{in}$ and mode temperature $T_{mode}$ \cite{AKMreview}: 
\begin{equation}
{\cal A}_0= {\cal M} \, P_{in}  T_{mode} \label{eq2}.
\end{equation}
This optomechanics scheme alters neither the measured position of the sideband peaks (detuned by $\pm \Omega_m[T_{beam}]$), nor their linewidth $\gamma_m(T_{beam})$: both are 
determined by
mechanical properties, which depend on the beam temperature $T_{beam}$.
The lineshapes are Lorentzian.
We introduce the {\it number of stored photons} in the cavity $n_{cav}$, function of both $P_{in}$ and $\Delta$:
\begin{equation}
n_{cav}\left(P_{in}\right) = \frac{P_{in} \, \kappa_{ext}/2}{\hbar \left( \omega_c + \Delta \right) \left(\Delta^2 + \kappa_{tot}^2/4 \right) } \label{eq0}.
\end{equation}

		\begin{figure}[h!]
		\centering
	\includegraphics[width=11cm]{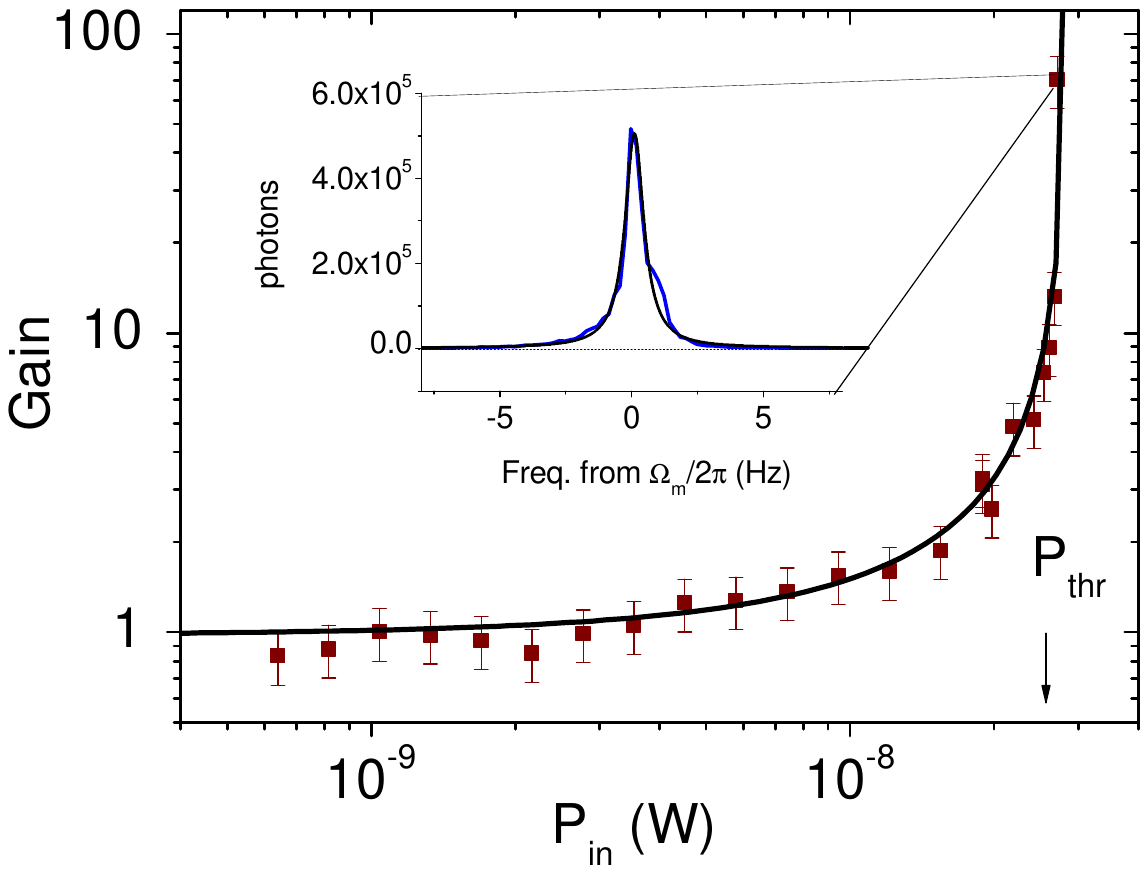}
	\vspace*{-2cm}
			\caption{
			Main: gain of the parametric amplification method based on the blue-detuned pumping scheme (210$~$mK data, brown squares), as a function of $P_{in}$. 
			As in Fig. \ref{fig_3}, the arrow indicates the position of the threshold $P_{thr}$ towards self-sustained oscillations.
			Inset: resonance line (blue trace) measured at very large gains, demonstrating its lorentzian lineshape (linewidth of order 0.9$~$Hz, close to instability). The black lines are fits, and we report about 20$~$dB amplification of the Brownian signal. Finite error from both statistics and fluctuations in mechanical parameters (see text). }
			\label{fig_4}
		\end{figure}

On the other hand for blue-detuned pumping, 
as we increase the injected power $P_{in}$ (but keep it below the instability threshold), the area $\cal A$ of the Stokes peak is amplified.
The blue/red-detuned pumping expressions write \cite{AKMreview}:
\begin{eqnarray}
\gamma_{ef\!f}\left(P_{in}\right) & = & \gamma_m - \mbox{Sign}\left(\Delta\right) \frac{4 g_0^2 \, n_{cav}\left(P_{in}\right) }{\kappa_{tot}} \label{eq1} , \\ 
{\cal A} & = & {\cal A}_0 \times \frac{\gamma_m}{\gamma_{ef\!f}\left(P_{in}\right)} \label{eqgain} ,
\end{eqnarray}
in the limit of negligible cavity thermal population.
For $\Delta>0$, the last term in Eq. (\ref{eqgain}) after the $\times$ sign is a gain, illustrated in Fig. \ref{fig_4}. It arises from the anti-damping, with $\gamma_{ef\!f}$ the linewidth of the Lorentzian peak.
Controlling the applied power $P_{in}$, from the knowledge of system parameters 
one can straightforwardly recalculate the value of ${\cal A}_0$, and thus of $T_{mode}$ (i.e. the temperature of the mode {\it in absence of} optomechanical pumping). In Fig. \ref{fig_4} we demonstrate 18.5$~$dB gain, which is greater than the previously reported maximum for a similar setup using a graphene device \cite{delftsteeleOscill}. 
Essentially {\it only} $\gamma_m$ depends on temperature, and has to be known to apply Eq. (\ref{eqgain}). It can be obtained easily from a measurement of the mechanical effective damping (the linewidth of the Lorentzian Stokes peak, Fig. \ref{fig_3}), by either extrapolating to $P_{in} \rightarrow 0$ or defining the position of the threshold $P_{thr} \propto \gamma_m$ [with Eq. (\ref{eq1}) at $\gamma_{ef\!f}=0$, see Figs. \ref{fig_3}, \ref{fig_4} and \ref{fig_oscill}]. More details are given in Appendix \ref{optomechs}.
Obviously, the main requirement for this $T_{mode}$ estimate is the {\it stability} of experimental parameters. The mechanical mode itself happens to be the limiting element (see Fig. \ref{fig_3} and Fig. \ref{fig_5} below), leading to finite error bars at large gains in Fig. \ref{fig_4}; fluctuations are further discussed in Sections \ref{inequil} and \ref{unstable}, see Fig. \ref{fig_8}.

		\begin{figure}
		\centering
	\includegraphics[width=11cm]{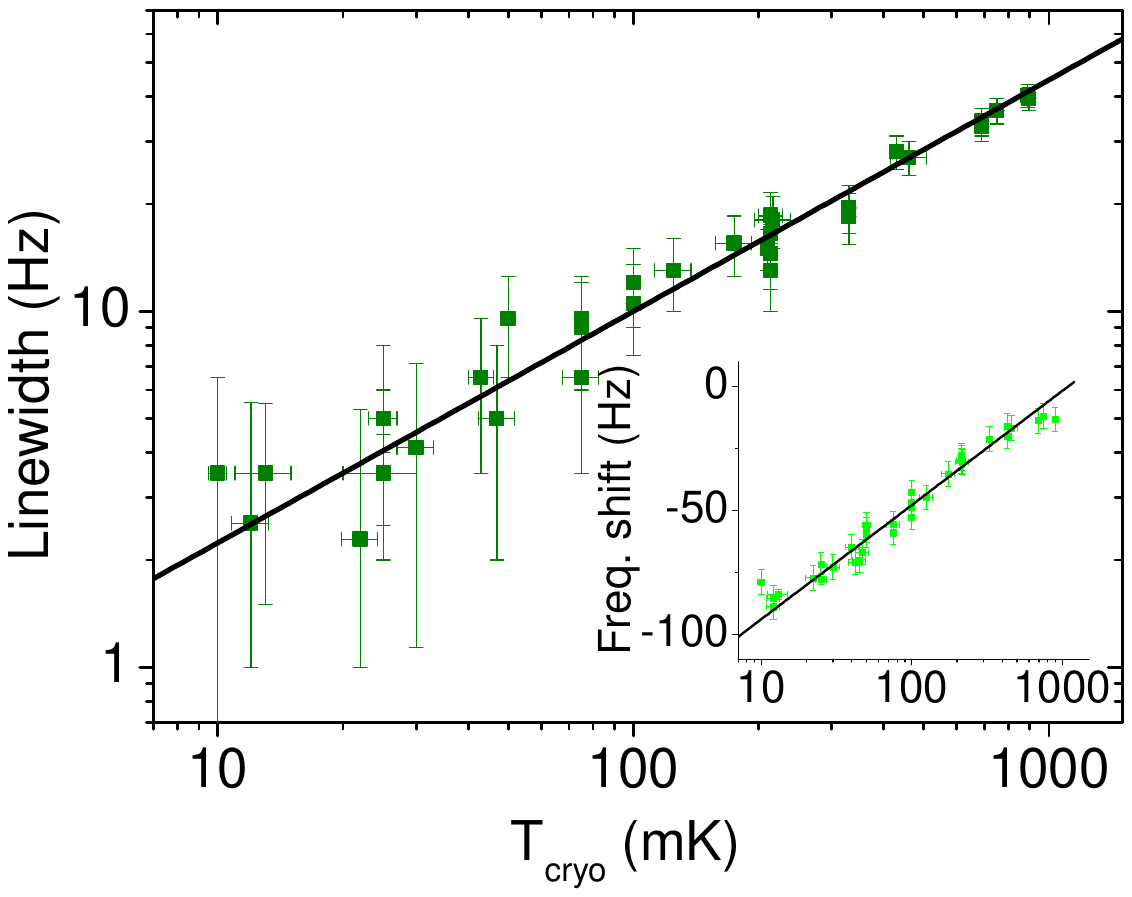}
	\vspace*{-2cm}
			\caption{
			Main: mechanical damping parameter $\gamma_m$ as a function of cryostat temperature $T_{cryo}$. Inset: resonance frequency shift from $\Omega_m(T^{\ast})$ of the flexural mode in same conditions. The black lines are fits following the TLS model ($\alpha \approx 0.65$, $T^{\ast} \approx 1.1~$K see text). Note the scatter in the data (also in Fig. \ref{fig_3}), further discussed in Section \ref{unstable}.}
			\label{fig_5}
		\end{figure}

The measured mechanical damping rates $\gamma_m$ and resonance frequencies $\Omega_m$ are shown in Fig. \ref{fig_5}.
The displayed dependencies are characteristic of NEMS devices in the millikelvin range: a damping $\gamma_m \propto T^\alpha$ with $0.3 < \alpha < 2.5$ 
and a logarithmic frequency shift $\propto \ln \left( T/T^{\ast} \right)$ with $T^{\ast}$ a characteristic temperature (see fits in Fig. \ref{fig_5}). 
For all materials (from monocrystalline to amorphous) this behavior is understood as a signature of TLSs (Two-Level Systems) \cite{TLSMohanty,TLSPashkin, KunalPRB,KunalPRL,FaveroTLSPRL,DavisTLS,Painter}: either defects (e.g. for monocrystalline Si, or polycrystalline Al), or constitutive of the atomic arrangement (for amorphous SiN). Direct coupling of the first flexural 
mode to the phonon bath (i.e. clamping losses) \cite{RoukesLifshitzClamp} is negligible for these structures at millikelvin temperatures.

Within the TLS model the mechanical mode is coupled to the two level systems, which are themselves coupled to the external bath: the electrons 
and the (thermal) phonons present in the moving structure \cite{KunalPRL}. For superconducting materials, the electronic contribution is negligible and the TLSs temperature should reflect the phononic temperature in the beam (i.e. the temperature of high frequency modes well-coupled to the clamping ends), 
 which we simply define as $T_{beam}$. By inverting the fits in Fig. \ref{fig_5} it is thus straightforward to extract $T_{beam}$.

		\begin{figure}[h!]
		\centering
	\includegraphics[width=11cm]{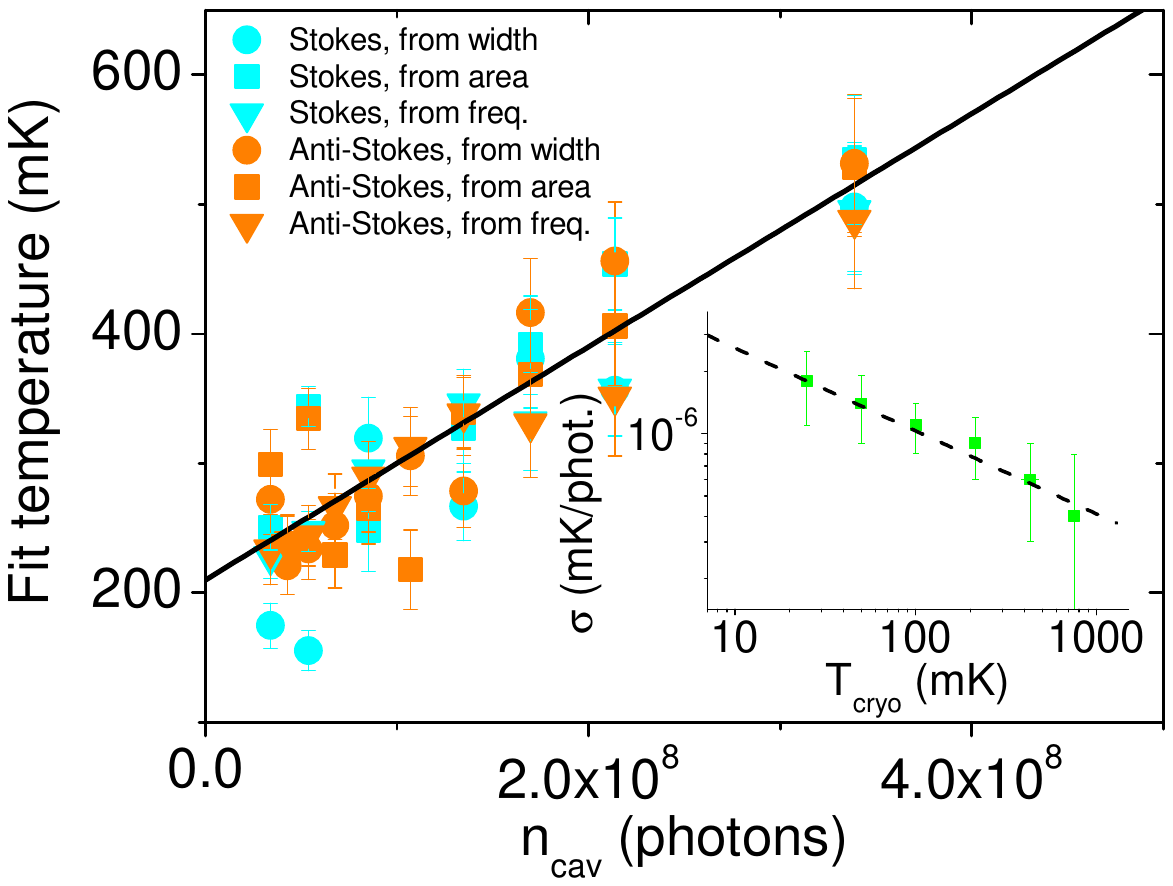}
	\vspace*{-2cm}
			\caption{
			Main: microwave heating as a function of $n_{cav}$ performed at 210$~$mK, with temperatures recalculated from the measured width and position (corresponding to $T_{beam}$) and peak area ($T_{mode}$). The line is a linear fit, leading to the microwave-heating coefficient $\sigma$. Inset: heating coefficient $\sigma$ versus cryostat temperature $T_{cryo}$; the dashed line is a power law guide to the eye. }
			\label{fig_6}
		\end{figure}

The aim of our work is thus to compare the temperature of the cryostat $T_{cryo}$ to $T_{beam}$ and $T_{mode}$. 
These results are analyzed in Section \ref{inequil}; however it is mandatory to quantify beforehand the impact of the microwave pump power on the defined temperatures. 
For this purpose we use the ``in-cavity'' pumping scheme (Fig. \ref{fig_2}).
We measure, at a given temperature $T_{cryo}$, the mechanical characteristics $\gamma_m$, $\Omega_m$ and the area ${\cal A}_0$ of the two sideband peaks as a function of injected microwave power $P_{in}$. Using respectively the fits of Fig. \ref{fig_5} and Eq. (\ref{eq2}), we can recalculate the expected temperatures $T_{beam}$ and $T_{mode}$ under microwave irradiation.
Since the local heating should be proportional to the local electric field squared confined onto the NEMS, we discuss these results as a function of $n_{cav}$.
The properties of the cavity itself as a function of the pump settings are discussed in Appendix \ref{optomechs}.

\begin{figure}
	\centering		 
			 \includegraphics[width=11cm]{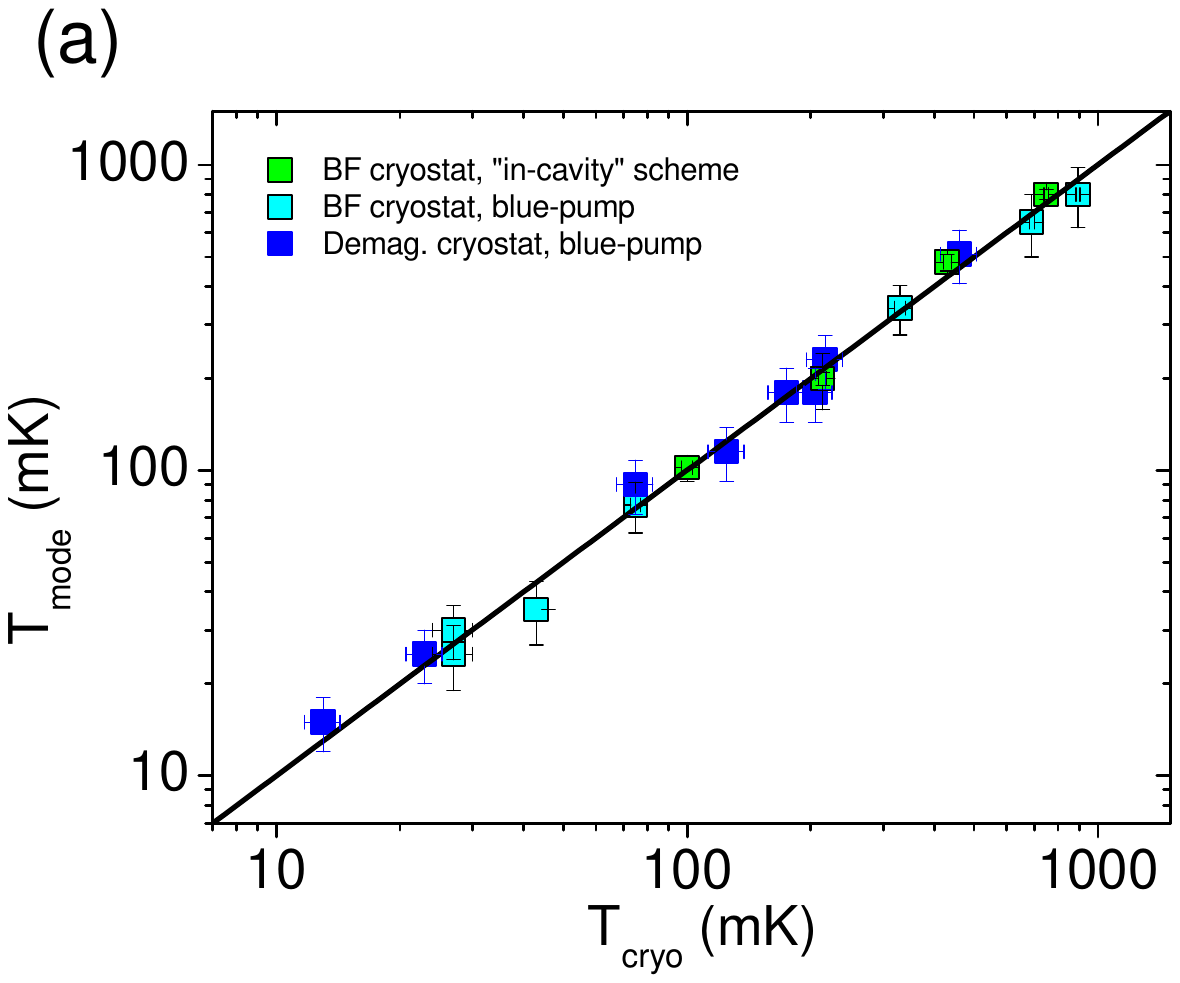}			 
			 \includegraphics[width=11cm]{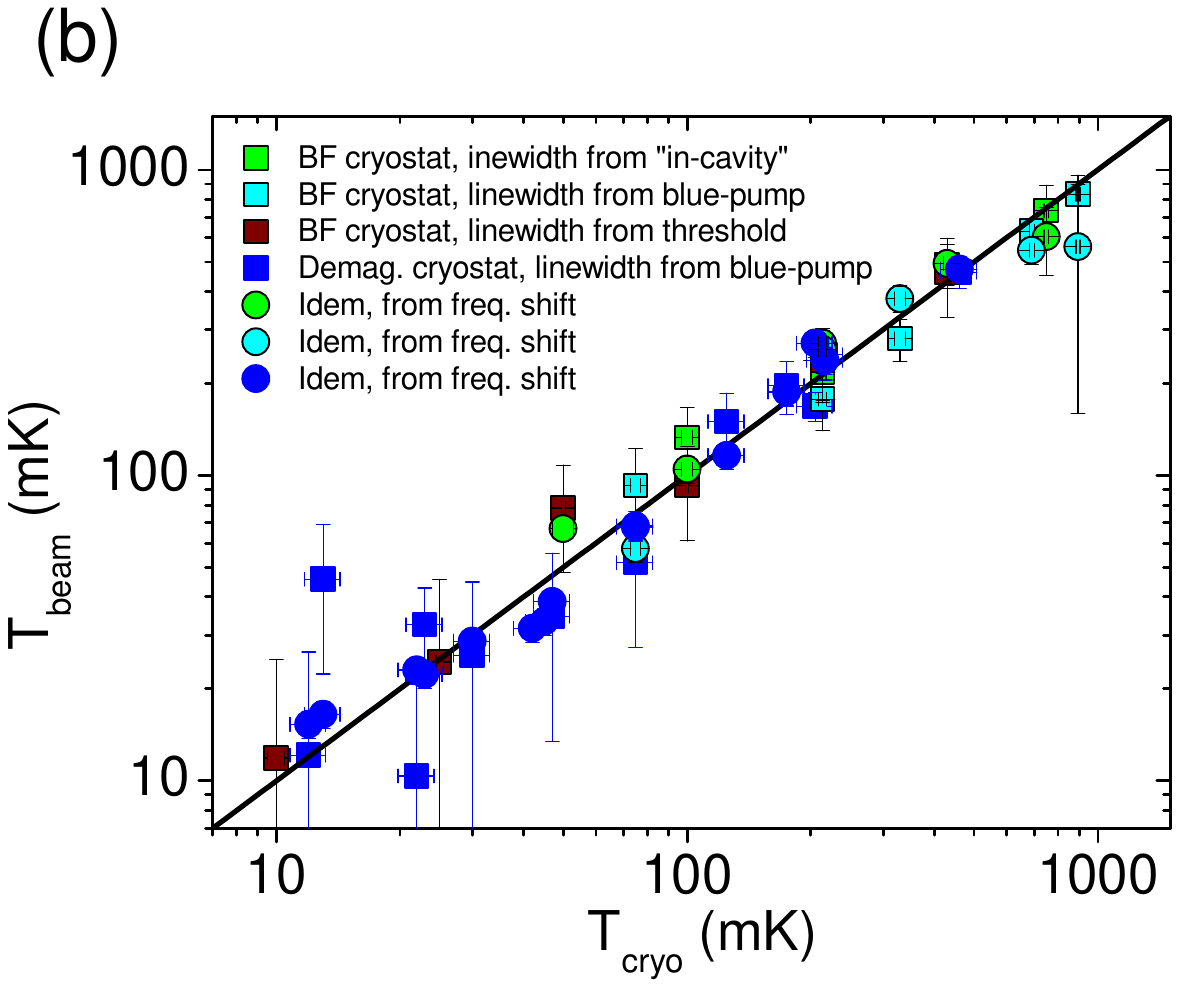}
			\vspace*{-2cm}
			\caption{ (a): mode temperature $T_{mode}$ as a function of cryostat temperature $T_{cryo}$. (b): beam temperature $T_{beam}$ inferred from the TLS bath as a function of $T_{cryo}$. The conditions of the measurements are defined in the legend. The line is the $y=x$ function. }
			\label{fig_7}
		\end{figure}

A typical result obtained at 210$~$mK is shown in Fig. \ref{fig_6} (main graph). Both $T_{beam}$ (obtained equivalently from damping and frequency shift) and $T_{mode}$ display {\it the same} linear dependence on $n_{cav}$, and the two sidebands are equivalent: this demonstrates that the effect is indeed thermal.
Defining the slope of the fit as $\sigma$, we can extract this coefficient as a function of $T_{cryo}$ (Fig. \ref{fig_6} inset).
This temperature-dependence is non-trivial, and no heating model is provided here: such a model should take into account the microwave absorption in the materials, the energy flow in the beam {\it plus} the clamping zone slab (suspended by the fabrication undercut), and finally the anchoring to the bulk of the chip.
Nonetheless, we can use this graph to estimate the NEMS heating for a given $T_{cryo}$ and $n_{cav}$ in the blue-detuned pumped scheme.
As a result, we extrapolate that applying a power of order $P_{thr}$ at 1$~$mK would heat the beam by about 1$~$mK; above 10$~$mK, the heating is essentially negligible (see error bars on $T$ axis of Fig. \ref{fig_oscill}).
Knowing the smallness of the coupling $g_0$ employed here, this demonstrates the capabilities of the method. 
Furthermore, because of this microwave-heating it is obviously meaningless to report experiments below about $T_{cryo} \approx 1~$mK for this first ``ultimate'' cooling attempt. 

\section{In-equilibrium Results}
\label{inequil}

From fits to Eq. (\ref{eqgain}) of the power-dependent Stokes peak area, we thus extract $T_{mode}$.
Reversing the fits of the mechanical parameters $\gamma_m$, $\Omega_m$ (Fig. \ref{fig_5}) we obtain $T_{beam}$.
Both are displayed as a function of $T_{cryo}$ in Fig. \ref{fig_7}, for our two experimental setups.
We demonstrate a thermalization from about 10$~$mK to 1$~$K of the mode and of the whole beam; the device {\it is in thermal equilibrium} over 2 orders of magnitude in $T_{cryo}$.
Reported lowest thermodynamic temperatures in the literature lie all within the range 10 - 30$~$mK \cite{cleland2010,quantelecsimmonds,quantelec2,lehnert2008,kippenbergamplifnonrecip,schwavsciencesqueeze,sillanpaaintrique,microwaveSBCoolLehnertPRL,delftsteeleOscill,jfink}; 
however one work reports a potential mode temperature for an Al-drum of order 7$~$mK, consistent with base temperature of dry dilution cryostats \cite{schwab7mK}.
Similarly, a lowest temperature of 7$~$mK is reported for a gigahertz phononic crystal \cite{Painter}; but obviously such a mode cannot be used for phonon thermometry at millikelvin temperatures.

\begin{figure}
		\centering
	\includegraphics[width=9.5cm]{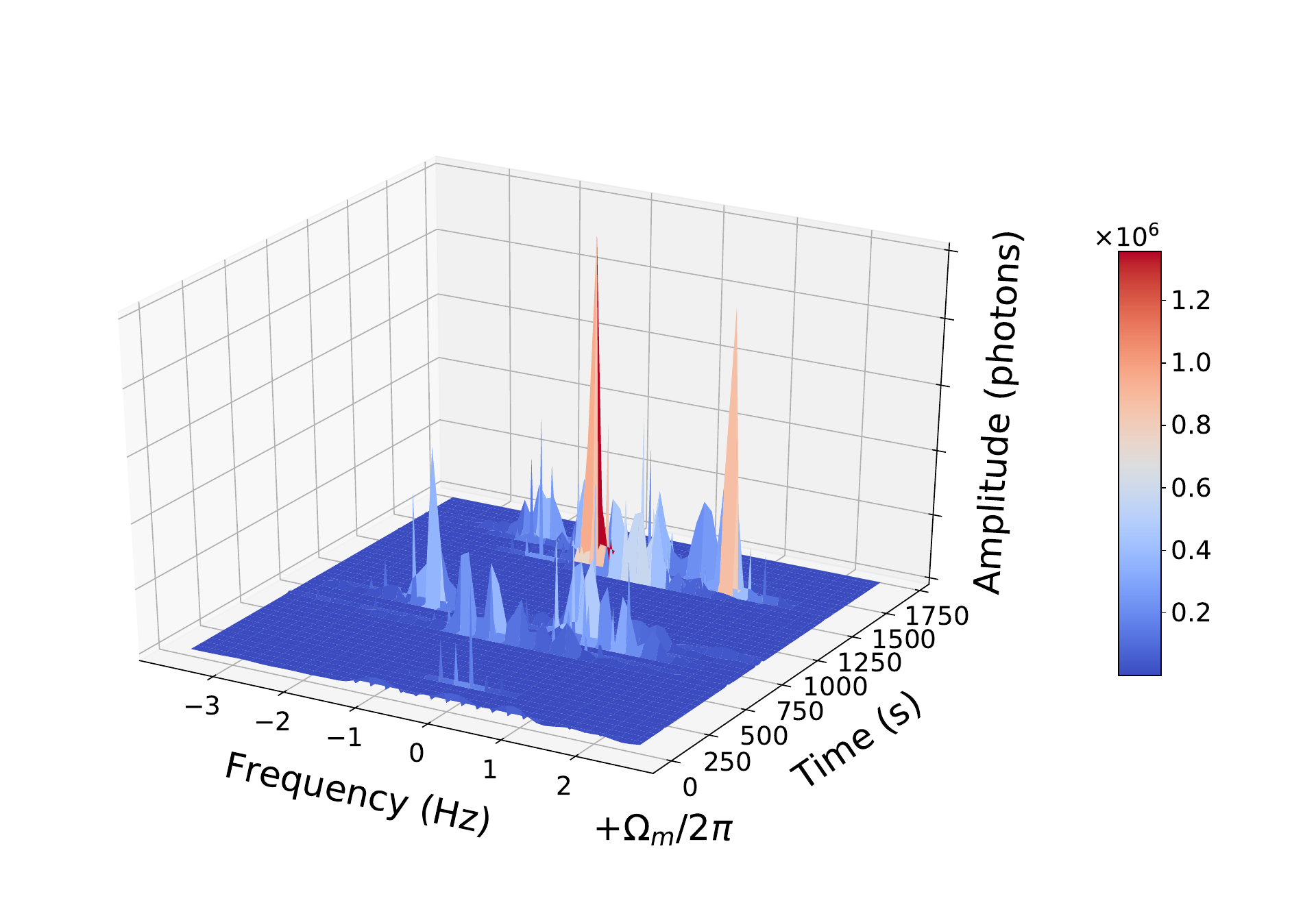}
			\caption{
			Stokes resonance peak (amplitude in color scale, frequency from $\Omega_m$ on the left) as a function of time at about 1$~$mK (
			0.6$~$nW applied power, blue-detuned pumping scheme). Huge amplitude jumps are seen, together with frequency (and damping) fluctuations (see text).}
			\label{fig_8} 
		\end{figure}

As far as thermalization between {\it measured} $T_{cryo}$ and $T_{mode}, T_{beam}$ is concerned, Fig. \ref{fig_7} reproduces the state-of-the-art, but does not go yet beyond even though the cryostat cools well below 7$~$mK. 
The reason for this is discussed in Section \ref{unstable}: below typically 100$~$mK, the system displays {\it huge amplitude fluctuations} which hinder the measurements (Fig. \ref{fig_8}).
These features have been seen by other groups for {\it beam-based microwave optomechanical devices containing an Aluminum layer}, but never reported so far \cite{sillanpaPrivCom,schwabPHD,contactLehnert,jfinkpriv}. 
Until recently, 
this essentially prevented experiments from being performed on these types of devices below physical temperatures of order 100$~$mK \cite{sillanpaamultimode,schwabrocheleau,TeufelBEAM}; 
 remarkably however, (Al covered) ladder-type Si beams \cite{jfink,jfinkpriv} seem to be less susceptible to this problem than simple doubly-clamped beams.
On the other hand, large signal fluctuations {\it have not} been observed to date 
 for (Al) drum-like structures 
\cite{sillanpaPrivCom,contactLehnert}, 
and do not show up in schemes which {\it do not} involve microwaves (e.g. magnetomotive measurements of SiN and Al beams \cite{OlivierPhD, TLSPashkin}, or laser-based measurements of Si beams \cite{DavisTLS}). 
This is what enabled nano-mechanical experiments to be conducted at base temperature of dilution cryostats.

Up to now, the only possibility to deal with these large events was {\it post-selection}, which is extremely time-consuming and even stops being useable at all at the lowest temperatures.
The origin of this phenomenon remains unknown, and we can only speculate on it in Section \ref{unstable} hereafter.
Note that there is {\it no evidence} of thermal decoupling in Fig. \ref{fig_7}.

More conventional frequency $\Omega_m$ and damping $\gamma_m$ fluctuations \cite{ACSnanoUs} are also present (see e.g. Fig. \ref{fig_8}). 
These features have been reported for essentially all micro/nano mechanical devices, as soon as they were looked for; 
their nature also remains unexplained, and their experimental magnitude is much greater than all theoretical expectations \cite{RoukesHentzNatNanotech}.
Frequency noise essentially leads to inhomogeneous broadening \cite{OliveNJP}. 
It {\it does not} alter the area $\cal A$ measurement, but does corrupt both frequency and linewidth estimates. 
This noise comes in with a $1/f$-type component \cite{RoukesHentzNatNanotech,ACSnanoUs}, plus {\it telegraph-like jumps} \cite{MartialPhD}.
It leads to the finite error bars in Fig. \ref{fig_5} inset; below 10$~$mK, the mechanical parameters $\Omega_m$ and $\gamma_m$ cannot be measured accurately.
Similar damping fluctuations \cite{ACSnanoUs} are more problematic, since the amplification gain Eq. (\ref{eqgain}) depends on $\gamma_m$.
The error bars of Fig. \ref{fig_5} (main graph) are essentially due to this; they translate into a finite error for the estimate of the gain, Fig. \ref{fig_4}, which itself limits the resolution on $T_{mode}$ (Fig. \ref{fig_7}, top). 

\section{unstable drive force features}
\label{unstable}

In Fig. \ref{fig_8} we show a typical series of spectral acquisitions as a function of time, around 1$~$mK. 
We see very large amplitude fluctuations which start to appear around 100$~$mK, and get worse for lower temperatures (regardless of the scheme used): the ``spikes'' grow even larger, but more importantly {\it their occurrence} increases.
We studied these events in the whole temperature range accessible to our experiment. Their statistics seems to be rather complex, and shall be the subject of another Article. Key features are summarized in this Section.
For blue-detuned pumping the spikes worsen as pump power increases, while for red-detuned pumping it is the opposite, suggesting that the effective damping of the mode plays an important role.
With in-cavity pumping, spiky features are also present at very low powers, but not at high powers when the NEMS physical temperature exceeds about 100$~$mK.
The recorded heights can be as large as equivalent mode temperatures in the Kelvin range.
Around $10-30~$mK, post-selection becomes impossible.

\begin{figure}[h!]
	\centering		 
			 \includegraphics[width=8cm]{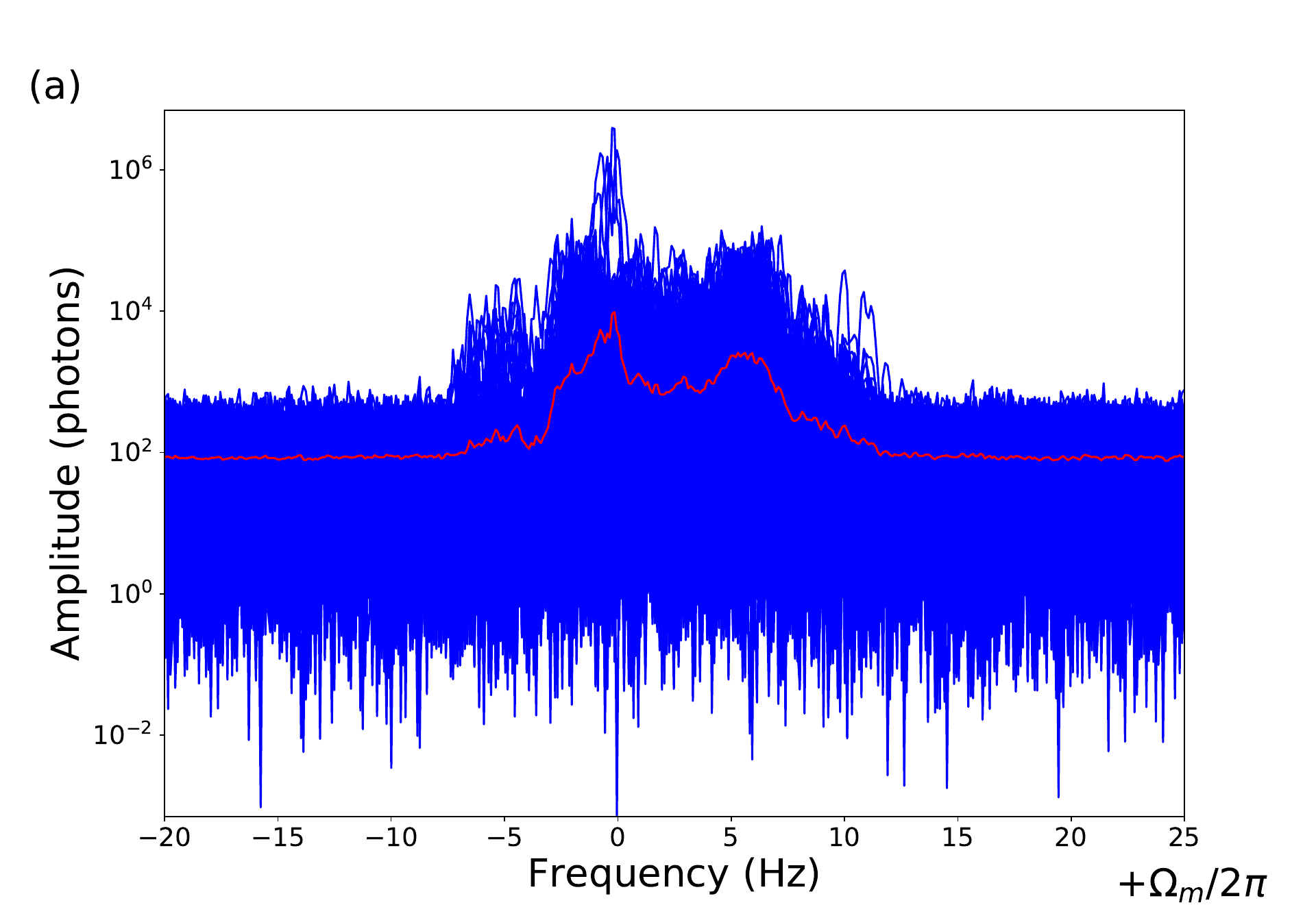}			 			 \includegraphics[width=8cm]{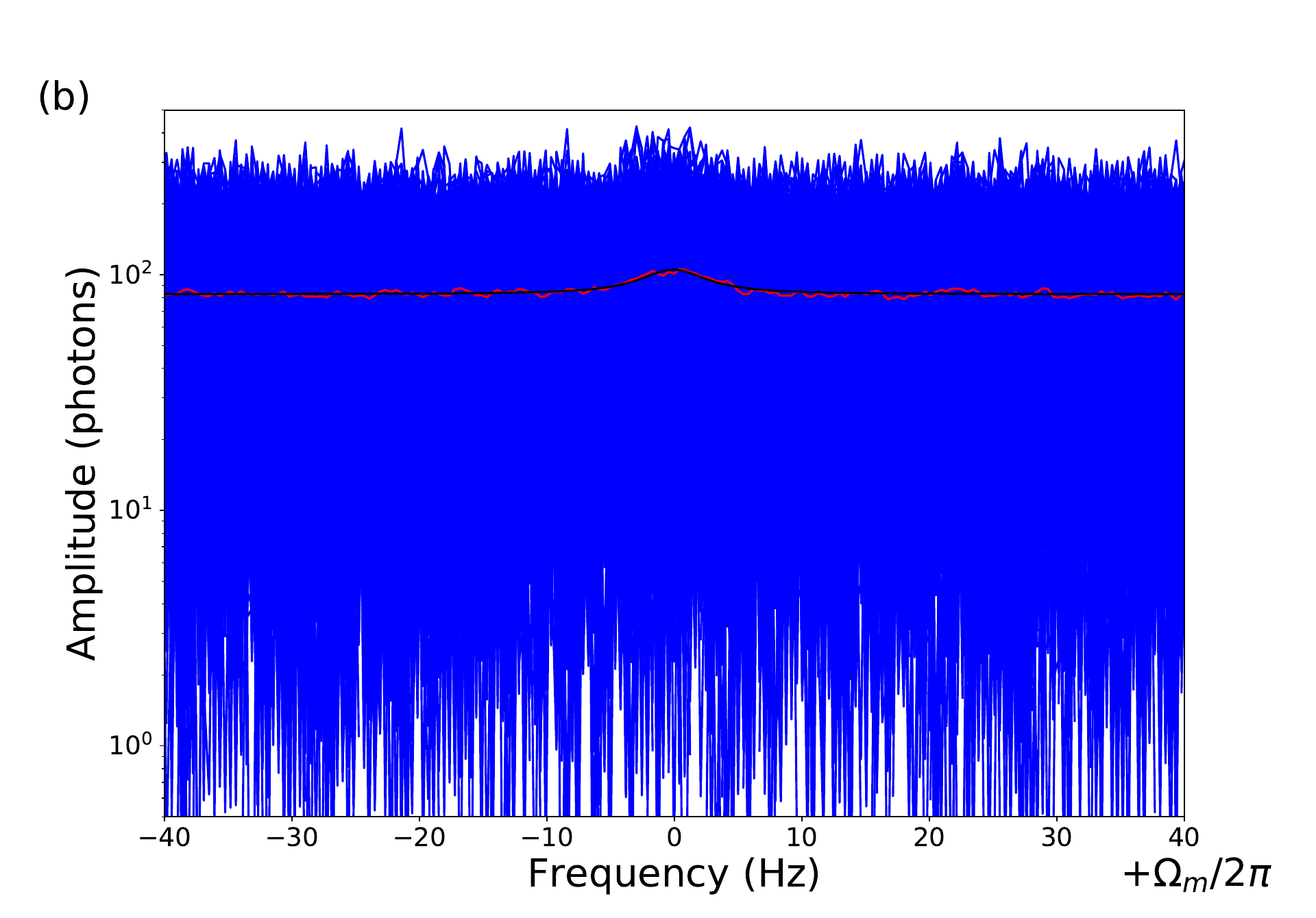}
			\caption{
			Comparing measurements with/without applied DC voltage in similar conditions. (a): resonance lines (blue) obtained at 5$~$mK with 
			0.6$~$nW drive; 2 hours acquisition shown in 560 averaged traces, with {\it no} DC voltage bias. (b): resonance lines obtained at 7$~$mK with 
			0.8$~$nW drive; 18 hours in 770 averaged traces, with +3$~$V DC applied on the transmission line (see text). The scheme used for both data sets is blue-detuned pumping, and the thick black line is a fit of the average curve (in red). Note the $10^4$ difference in the vertical axes. }
			\label{fig_9}
		\end{figure}
	
With the aim of searching for the origin of this effect, we have characterized it in various situations.
We first realized that cycling the system from the lowest temperatures to above 100$~$mK was producing a sort of ``reset''. But very quickly (a matter of hours), after cooling down again the large spikes happen to dominate the signal again. 
We then tried to apply a small magnetic field to the system; this was not very conclusive.
However, applying a DC voltage had a drastic effect on these random features. This is illustrated in Fig. \ref{fig_9}: with a few volts on the chip's coplanar transmission line {\it all the large features disappear}. The averaged signal (deeply buried into the noise) recovers a reasonable Lorentzian lineshape (see fit in Fig. \ref{fig_9}), while the shape of the spikes is not resolved (Fig. \ref{fig_8}).
Details on the DC voltage biasing are given in Appendices \ref{setups} and \ref{optomechs}.
	
In discussing the source of this feature, a few comments have to be made.
What is shown in Fig. \ref{fig_8} is primarily fluctuations of the {\it output} optical field.
These are detected {\it only} on the Stokes and Anti-Stokes peaks, for any of the schemes shown in Fig. \ref{fig_2}.
Furthermore, the threshold to self-oscillation in the blue-detuned pumping scheme displays a large hysteresis (certainly due to nonlinearities in the system, see Fig. \ref{fig_oscill}, Appendix \ref{optomechs}).
We have noticed that when the microwave power applied (at frequency $\omega_c+\Omega_m$) lies within this hysteresis, the spiky events seem to be able to trigger the self-oscillation. 
This would not be possible if the amplitude fluctuations measured were only in the detected signal, at the level of the HEMT.
We thus have to conclude that we see genuine {\it mechanical} amplitude fluctuations. 
However, these cannot be due to damping fluctuations alone that could trigger self-oscillations, since we do see the same type of features when pumping red-detuned or in-cavity.

If these fluctuations were due to the input field itself, from Fig. \ref{fig_6} we would reasonably conclude that the NEMS beam would be heated to rather high temperatures, leading to broad and very shifted in frequency (see Fig. \ref{fig_5}) Stokes/Anti-Stokes peaks. This is not compatible with the measurement of Fig. \ref{fig_8}.
The only reasonable conclusion seems thus to be that we do suffer from a {\it genuine extra stochastic force} acting on the mechanical element.
This is consistent with a stronger sensitivity to the phenomenon when the effective damping of the mode is small (bue-detuned scheme).
Since a DC voltage applied {\it only to the cell} can drastically modify the measured features, the source has to be on-chip.
Noting furthermore than with an in-cavity pumping, it disappears when the beam temperature exceeds 100$~$mK, we conclude that it {\it should even be within the mechanical element}.
But the mechanism remains mysterious: citing only documented effects in other areas of research, is it linked to vortex motion in the superconductor \cite{vortexNoise}, trapped charges \cite{chargeNoise}, adsorbed molecules \cite{adsorbate} or to atomic-size Two-Level-Systems in dielectrics (beyond the standard friction model) \cite{dielecTLS}? 

The low temperature properties of NEMS are described within the tunneling model of Two-Level-Systems (TLSs): for damping, frequency shifts, and phase fluctuations \cite{DavisTLS,fongPRB,ACSnanoUs}. It is thus natural to consider strongly coupled individual TLSs as the most probable source of our problems. Besides, while the actual nature of these microscopic defects remains elusive in most systems, 
they could be generated in many ways beyond the standard atomic configuration argument \cite{ustinov}; an electron tunneling between nearby traps would be a TLS strongly coupled to its electromagnetic environment, among other possibilities \cite{adsorbate}.
For Al-based NEMS, these would create (only a few) defects present in (or {\it on}) the Al layer; they should carry a dipole moment, which couples them to the microwave drive as well as to the electric field generated by the applied DC voltage. This field distorts their potentials, such that they could get locked in one state and ``freeze''.
Furthermore, our results seem to be very similar to those of Ref. \cite{ParpiaGamma} obtained with a {\it macroscopic} mechanical glass sample, where ``spiky'' events were demonstrated to be originating in the interaction with low-level radioactivity (gamma rays). These results suggest a parametric coupling to TLS at GigaHertz frequencies mediated by the microwave drive, but were the energy corresponding to the large peaks would be provided by the external radiation.

The reason why the mechanism should be dependent on the low phononic dimensionality or size of the device (typ. width about 100$~$nm, much smaller than the phonon wavelength at 10$~$mK) is nontrivial. 
One simple argument could be that the spring constant of the modes under consideration are very different: about 1$~$N/m for megahertz beams and 100$~$N/m for drumheads. This could justify why beam-based structures are more reactive to external force fluctuations; an immediate consequence of this argument is then that membrane-based Al devices {\it are not} truly immune to force fluctuations, but are just less sensitive: if that should be true, cooling them to low enough temperatures would eventually revive the same features as for beam-based NEMS.

		\begin{figure}
		\centering
	\includegraphics[width=12 cm]{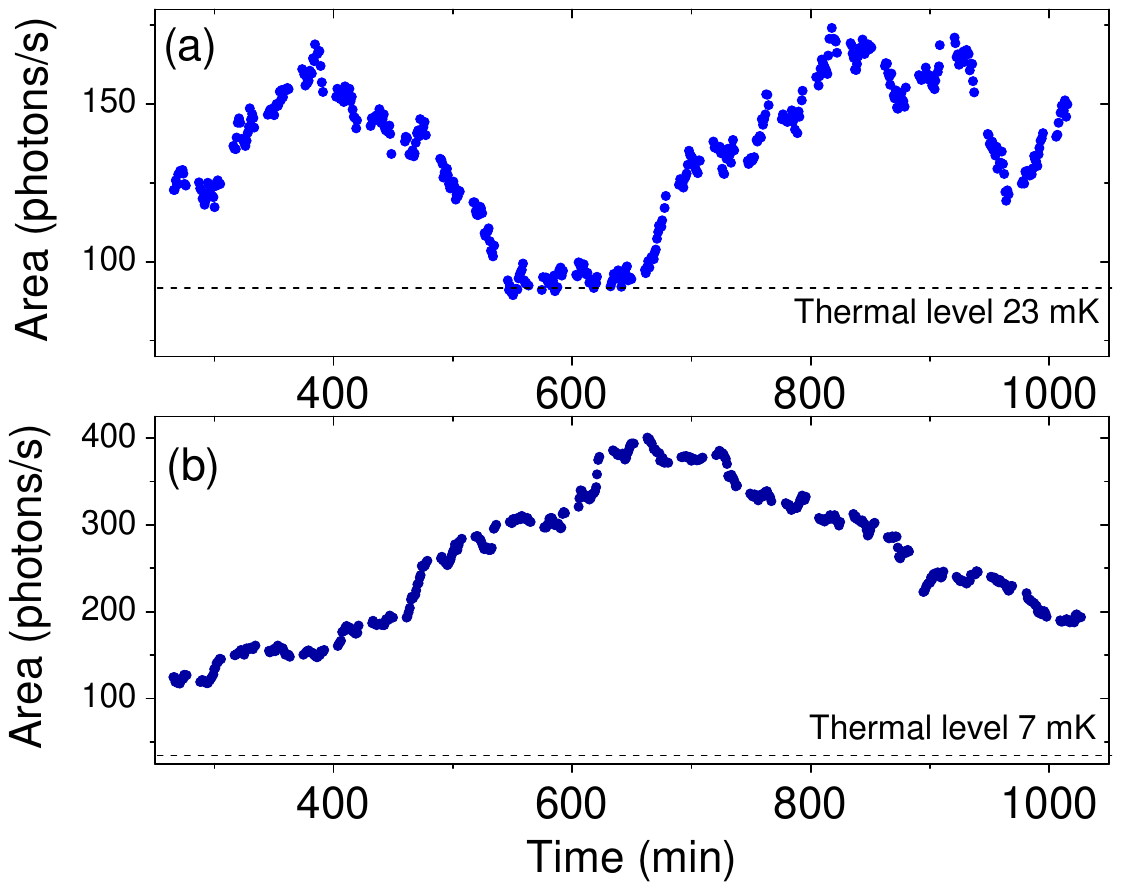}
	\vspace*{-2cm}
			\caption{
			(a): area of peak extracted from a sliding average performed with a window of 26$~$minutes at 23$~$mK, with applied power 
			0.8$~$nW (blue-detuned pumping). (b): same measurement performed at 7$~$mK. The horizontal dashed lines are the thermal population expected values, matched at 23$~$mK in the stable zone (middle of graph). At the lowest temperatures, we observe very large amplitude fluctuations which {\it cannot be} of thermal origin; the measured area remains always larger than the expected value (see text).}
			\label{fig_10} 
		\end{figure}
		
To conclude, let us concentrate on the measurements performed at ultra-low temperatures with a DC voltage bias (of the type of Fig. \ref{fig_9}, bottom). Even with the help of the in-built parametric amplification, the signal is very small and requires decent averaging, typically here about 30$~$minutes for reasonable error bars. 
Even if the resonance peak is Lorentzian, below typically 20$~$mK the measured area $\cal A$ does not correspond to the actual cryostat's temperature $T_{cryo}$: it is always bigger, but the actual value presents {\it large fluctuations} in magnitude.
This is demonstrated in Fig. \ref{fig_10}, with identical measurements performed at 23$~$mK and 7$~$mK. 
What is shown is how the measured area of the Stokes peak evolves over time, performing a sliding average over the whole set of acquired data.
In the former case, we see that the fluctuations of the measured area are not more than about $+ 60~\%$; they are much smaller for higher temperatures, leading to proper estimates of $T_{mode}$. However, for the latter these are greater than $300~\%$. Besides, fluctuations happen to have an extremely slow dynamics: while spikes switch on/off faster than our acquisition time, their overall occurrence fluctuates over {\it a day} (Fig. \ref{fig_10}).
By no means could this behavior be explained by a thermal decoupling of the device from the cryostat.
As a consequence, even the calm zones in Fig. \ref{fig_8} are corrupted by the phenomenon shown in Fig. \ref{fig_10}. This is essentially why no reliable data could be acquired below 10$~$mK;  
but from the DC biasing and the continuous monitoring of the Stokes peak, thermal equilibrium has been demonstrated at about ten times lower temperature than previously reported in microwave doubly-clamped NEMS experiments \cite{sillanpaamultimode,schwabrocheleau,TeufelBEAM}. 

\section{Conclusion}

We presented measurements of a microwave optomechanical system performed on a nuclear adiabatic demagnetization cryostat, able to reach temperatures well below the 10$~$mK limit of conventional dilution machines.
Relying on a fairly standard microwave wiring and the in-built parametric amplification provided by a blue-detuned pumping, we devised a method 
providing accurate thermometry of both the mechanical mode and its on-chip environment (the Two-Level Systems to which it couples).
The experiment was conducted on a beam Nano-Electro-Mechanical System embedded in an on-chip microwave cavity. The efficiency of the method is demonstrated with a very low opto-mechanical coupling.
Thermalization is shown from 10$~$mK to 1$~$K with no sign of thermal decoupling. 

However at very low temperatures we report strong fluctuations in the signal amplitude which prevented any experiments to be conducted at ultra-low temperatures. These features appear around 100$~$mK and have been observed in different laboratories, but had never been studied in details so far.
We demonstrate the characteristics of these fluctuations, and argue that they are due to an extra stochastic driving force of unknown origin.
Microwave irradiation seems to trigger the phenomenon.
Applying a DC voltage of a few Volts on-chip cancels the large spiky events, but a small component of this extra random drive persists, with variations over a typical timescale of about a day.

It is unclear if all the fluctuations characteristics present in the devices (amplitude, frequency, damping) are linked to the same underlying mechanism.
One could even imagine that temperature-dependent non-linear effects could impact the phonon-photon coupling, beyond the lowest (geometrical) order $g_0$.
However, it is likely that these effects are present in {\it all} experimental systems at different levels of expressions, since all NEMS/MEMS share the same overall characteristics (especially damping, frequency shifts and phase noise typical of Two-Level Systems physics).
It is thus tempting to relate this stochastic force to a mechanism mediated by some kind of microscopic TLSs, driven by microwaves but blocked under DC voltage biasing.
This stochastic driving force can mimic to some extent a thermal decoupling, and could explain why some drumhead devices in the literature refuse to cool down below typically 20-30$~$mK. 
While being a limitation for experimentalists, this phenomenon definitely deserves theoretical investigations. 
The present work also calls for further experiments at lower temperatures, using other types of devices (e.g. drums). This shall be performed in the framework of the European Microkelvin Platform (EMP) \cite{EnssPickettEMP}.

{\it Note added in proof:} following the present work, a collaboration between {\it Aalto University} and {\it Institut N\'eel} has started with the aim of cooling down a drumhead Al device as low as possible on our adiabatic nuclear demagnetization platform. We have evidence that the same features as for beams are present, at a different level of expression. This shall be published elsewhere.

(\dag) Corresponding Author: eddy.collin@neel.cnrs.fr

\begin{acknowledgements}

We acknowledge the use of the N\'eel facility {\it Nanofab} for the device fabrication, and the N\'eel {\it Cryogenics} facility with especially Anne Gerardin for realization of mechanical elements of the demagnetization cryostat. 
X.Z. and E.C. would like to thank Olivier Arcizet and Benjamin Pigeau for help in room-temperature characterization of the NEMS devices using optics; E.C. would like also to thank O. Arcizet, M. Sillanp\"a\"a, K. Lehnert, J. Teufel, S. Barzanjeh, K. Schwab, J. Parpia and M. Dykman for very useful discussions.
We acknowledge support from the ERC CoG grant ULT-NEMS No. 647917, StG grant UNIGLASS No. 714692 and the STaRS-MOC project from {\it R\'egion Hauts-de-France}. 
The research leading to these results has received funding from the European Union's Horizon 2020 Research and Innovation Programme, under grant agreement No. 824109, the European Microkelvin Platform (EMP).

\end{acknowledgements}


\appendix

		\begin{figure}
		\centering
	\includegraphics[width=9cm]{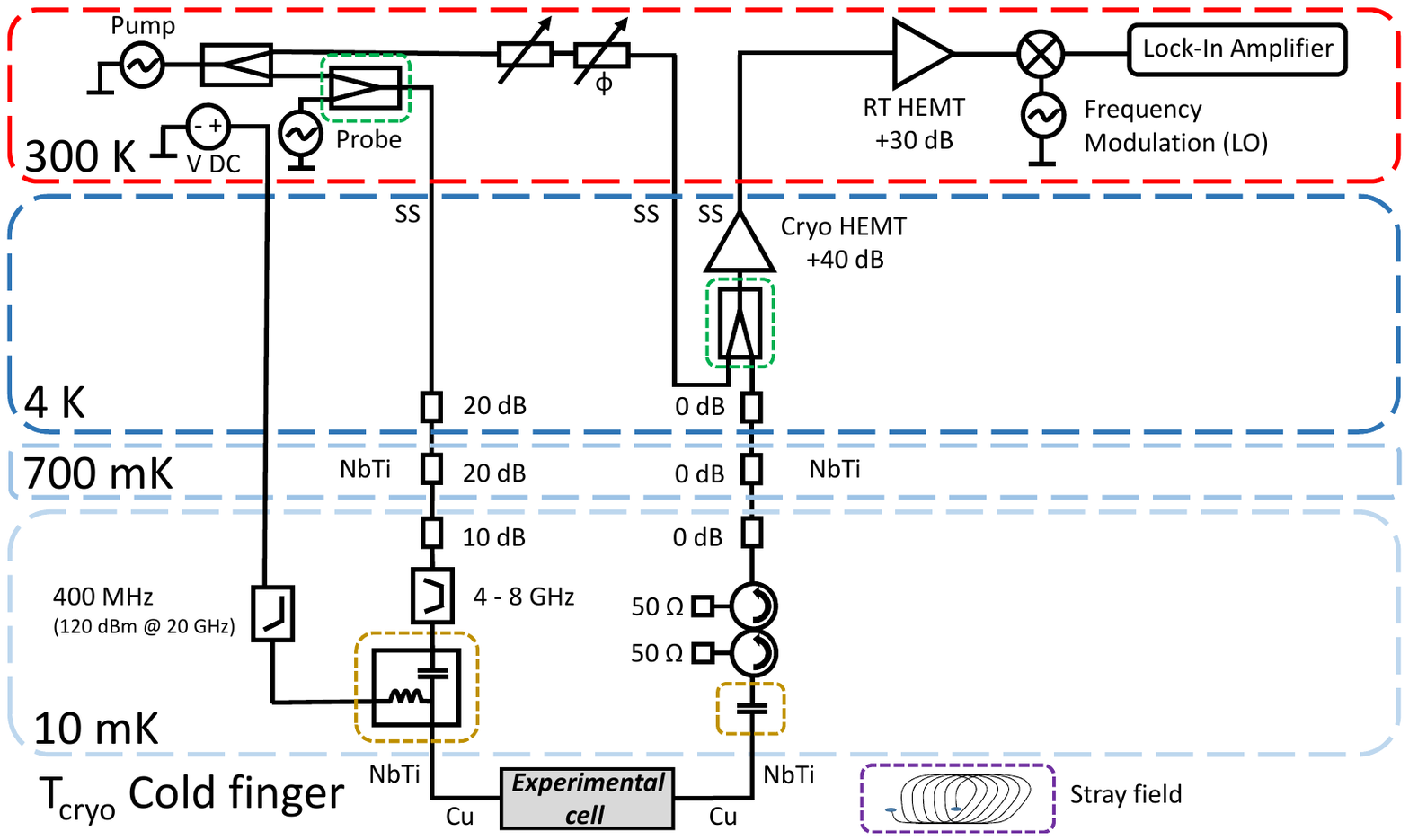}
	\vspace*{-0.75cm}
			\caption{
			Simplified common wiring of the experimental platforms; the different levels within the cryostats (BF and demag.) are shown with their respective temperature. SS stands for Stainless-Steel, NbTi for Niobium-Titanium and Cu for Copper coaxial cables (50$~$Ohms impedance). The boxed elements have been added/removed depending on the experimental run (see text for details).}
			\label{fig_wiring}
		\end{figure}

\section{setup}
\label{setups}

Two similar microwave setups have been used in these experiments.
Their common features are described in Fig. \ref{fig_wiring}. This wiring is basic and can also be found within the literature \cite{lehnert2008}.
Essentially, they are built around a cryogenic HEMT (High Electron Mobility Transistor) placed at about 4$~$K and two circulators mounted on the mixing chamber of the dilution units. 
On the BlueFors$^{\textregistered}\,$ (BF) machine, the HEMT is a Low Noise Factory$^{\textregistered}\,$ 4-8$~$GHz bandwidth, with a measured noise of about 3$~$K (10 photons) at 6$~$GHz. On the nuclear adiabatic demagnetization cryostat, it is a Caltech 1-12$~$GHz bandwidth with a measured noise of about 15$~$K (50 photons at 6$~$GHz).
The (dashed green) boxed component below the HEMT in Fig. \ref{fig_wiring} represents a power combiner used to realize an opposition line. On the BF setup, this is mandatory to avoid saturation of the cryogenic HEMT from the strong blue-detuned pump tone. On the demag. cryostat, the cryogenic HEMT is linear enough so this protection is not necessary. 
This choice has been made because of space constrains: feeding an extra microwave opposition line in the nuclear adiabatic demagnetization cryostat would be very demanding.
The filtering of the injection lines (DC and microwave) is also described in Fig. \ref{fig_wiring}.

Gains and noise levels of the full chain have been carefully checked with respect to HEMT working point. 
Besides, each component has been tested at 4$~$K prior to mounting. The whole setup has then been calibrated, using an Agilent$^{\textregistered}\,$ microwave generator EXG N5173B and an Agilent$^{\textregistered}\,$ spectrum analyser MXA N9020A. 
The measurements presented in the core of the paper have been realized using a Zurich Instruments$^{\textregistered}\,$ UHFLI lock-in detector operating in spectrum mode. The signal is mixed down with a Local Oscillator (LO) and detected at frequency $\pm \Omega_m+2~$MHz (the shift avoiding overlap of Stokes/Anti-Stokes signals). 
The generators used were from Agilent$^{\textregistered}\,$ 
or Keysight$^{\textregistered}\,$ brands leading to equivalent data quality. 
The {\it absolute} error in the calibrations is estimated at about $\pm 2~$dB over the whole set of realized runs; within these error bars, the two cryogenic platforms gave the same quantitative results.

Particular care has been taken in thermalization issues.
The experimental cell is thus made of annealed high-purity copper. It is mounted on a cold finger either bolted onto the mixing chamber plate of the BF machine, or connected through silver wires to the bottom of the nuclear stage of the demag. cryostat. 
This stage is of laminar type, about 1$~$kg of high-quality copper \cite{paperYuriSatge}. On the top it is connected through a Lancaster-made Al heat switch \cite{ULANCHSpaper} to the mixing chamber of the home-made dilution unit. Both dilution cryostats reach base temperatures of order 7-10$~$mK depending on cooling power settings.
The PCB board mounted inside the cell is hollow in its center; the chip is pressed there {\it directly} onto the copper block, by means of a copper/indium spring.
At very large microwave powers, we do see heating at the intermediate stages where attenuators are anchored. However, the nuclear stage thermometers {\it do not} show any heating, even at the highest powers used.

	\begin{figure}[h!]
		\centering
	\includegraphics[width=11.5cm]{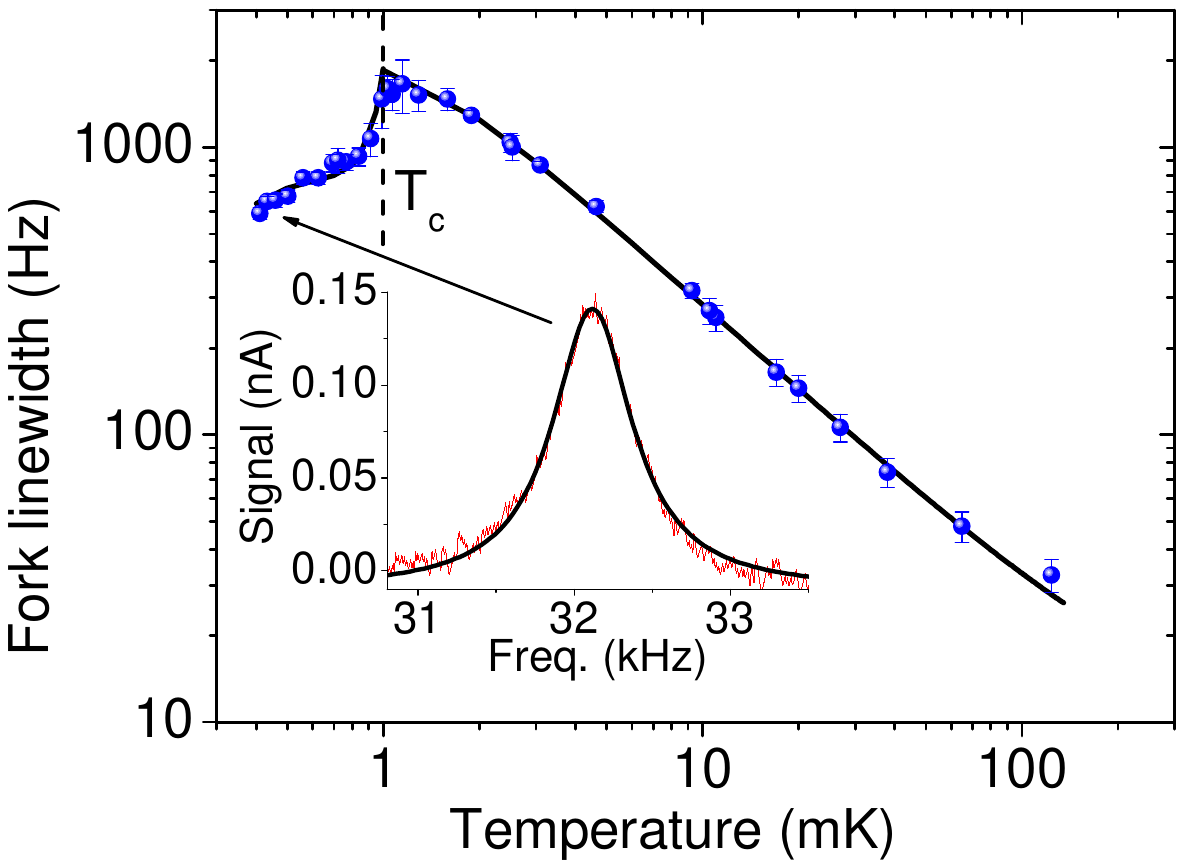}
	\vspace*{-2cm}
			\caption{
			Calibration measurements performed on the $^3$He thermometer, from the lowest operated temperature (here 400$~\mu$K, raw data fork resonance curve in inset with Lorentzian fit) to about 100$~$mK. The superfluid transition $T_c$ is clearly visible around 1$~$mK and can be used as a temperature fixed point; the line is the expected behavior from $^3$He viscosity, see text.}
			\label{fig_3He}
		\end{figure}

Thermometry below typically 30$~$mK is no easy task; below 10$~$mK it requires the expertise of ultra-low temperature laboratories (see e.g. EMP partners \cite{EnssPickettEMP}). In the graphs of Fig. \ref{fig_7}, the $x$ axis is as important as the $y$ one; this point is not always emphasized in the literature. We thus paid particular care in providing an almost primary temperature scale for our two microwave platforms.
The use of the word {\it almost} here should be understood as follows: we also used resistive thermometers (RuO$_2$, carbon Speer-type) calibrated against  primary devices, and our primary thermometers (which do follow known temperature dependencies) are calibrated at a single point in temperature for practical reasons. Our laboratory has a long history of working for the construction of the ultra-low temperature scale \cite{RefHenriULTS}.
In our case, the BF cryostat has been equipped with a Magnicon$^{\textregistered}\,$ MFFT noise thermometer and a CMN paramagnetic salt (Cerium Magnesium Nitrate) thermometer. The MFFT was bolted at mixing chamber level while the CMN was directly mounted on the cold finger.
On the demag. cryostat, an MFFT is mounted at the top of the nuclear stage while a $^3$He thermometer is connected at the bottom; the exerimental cell is actually {\it between} the nuclear stage and this thermometer.
When their working range overlap, the thermometers agree within typically 2 - 5 $~\%$; the error bars on $x$ axes of the temperature plots also take into account thermal gradients and slow drifts.

The nuclear adiabatic demagnetization process requires large magnetic fields to be used (here, 8$~$T). 
One usually starts at $T_{ini} \approx 10~$mK with the superconducting coil surrounding the stage fully magnetized at $B_{ini}$ (heat switch open), and then demagnetizes to about $B_{fin} \approx 100~$mT to reach the lowest possible temperatures (in the present runs, we worked down to 400$~\mu$K, Fig. \ref{fig_3He}). The cryostat can then stay cold over a week (heat leak $< 100~$pW); another cycle needs then to be initiated with a pre-cooling of the system. The process being adiabatic, the nuclear spins of copper are cooled according to $B_{ini}/B_{fin}=T_{ini}/T_{fin}$ \cite{PobellBook}.
Our magnet is compensated on both sides, but nonetheless small stray fields are present when it is magnetized (represented in Fig. \ref{fig_wiring} by the small boxed coil). In order to make sure that these fields do not introduce any slant in the data, we performed measurements {\it above} 10$~$mK but within a demagnetization cycle: we started at 20$~$mK insead of 10$~$mK, and reduced the field by 1/2. This can then be compared to experiments performed without any field, at base temperature of the dilution unit.
The microwave cavity does shift in frequency with applied field (see Appendix \ref{optomechs}), but no other difference could be found.
All data taken below 10$~$mK have been obtained exclusively with the nuclear adiabatic demagnetization technique.

The lowest part of our temperature scale is obtained from a $^3$He thermometer (see Fig. \ref{fig_3He}).
It is based on the viscosity measured in the fluid with an immersed probe, a mechanical resonator.
In the past, vibrating wires were the best choice for this task \cite{GuenaultViW,ClemensViW}. Today, people use quartz tuning forks which are more practical \cite{Rob3Hefork,ULANCFork,ULANCFork2}.
Our thermometer is thus a nested cell containing two tuning forks (one in the outside cell, the other in the inside), filled with silver sinters connected to the nuclear stage by silver wires. The outside cell serves as a thermal shield for the inner one, and does not cool down below typically 1$~$mK. The inner cell is directly connected to the cold finger that hosts the microwave cell.

$^3$He is a Fermi liquid above about 1$~$mK, and becomes superfluid below. At $0~$bar pressure, $0~$T field this state is called $^3$He-B, and the properties of this amazing fluid have been extensively studied over the years \cite{Book3HeVollhardt}.
For instance, the viscosity of $^3$He is well known for both normal and superfluid states, see e.g. \cite{HookHall1,HookHall2}.
In principle, from the knowledge of the geometry of the immersed object and its surface rugosity, one can calculate the friction and thus the broadening of  
the resonance. In practice, it is much more efficient to calibrate the device by scaling its properties on known measurements \cite{ULANCFork,ULANCFork2}.

In order to perform this scaling, we measured the damping of the quartz tuning fork thermometer (inner cell) as a function of the final field of the demagnetization process $B_{fin}$. The protocol was to demagnetize by steps, and then remagnetize to $B_{ini}$ in order to verify that the process was indeed adiabatic (by recovering the initial temperature $T_{ini}$). It is then straightforward to calculate the temperature of the nuclear spins of copper for each step, which should be in equilibrium with electrons (and finally $^3$He) if one waits long enough. 
The result is shown in Fig. \ref{fig_3He}. The line is the expected behavior from published results. Note that the superfluid transition $T_c$ also acts as a {\it fixed point} in the temperature scale. 

\section{Microwave optomechanics}
\label{optomechs}

The microwave cavity is the central element of the technique.
It happens to be very sensitive to external conditions: cell temperature, pump settings (power, detuning), DC voltage bias and magnetic field.
In order to avoid any systematic error, the protocol is thus to measure the cavity {\it for each setting}.
On the BF cryostat, in order to mimic stray fields from the demagnetization protocol and to study the impact of (small) magnetic fields on the unstable features reported in Section \ref{unstable}, a small coil has been installed on top of the cell (boxed element in Fig. \ref{fig_wiring}). It can create perpendicular fields up to about 10$~\mu$T.
The voltage DC bias is applied {\it only} to the cell which is placed within a Bias Tee and a DC Block (see boxed elements in Fig. \ref{fig_wiring}). Above about 4$~$V, the chip starts to heat (certainly because of currents flowing within the silicon). 
All reported measurements obtained with a DC bias have thus been acquired at 3$~$V DC, ramping slowly periodically (every 1/2 hour approximately) the voltage to almost 4$~$V and back again. At zero bias, after some time the ``spikes'' eventually reappear. They did not seem to react particularly to the small magnetic fields used.

		\begin{figure}[h!]
		\centering
	\includegraphics[width=11.5cm]{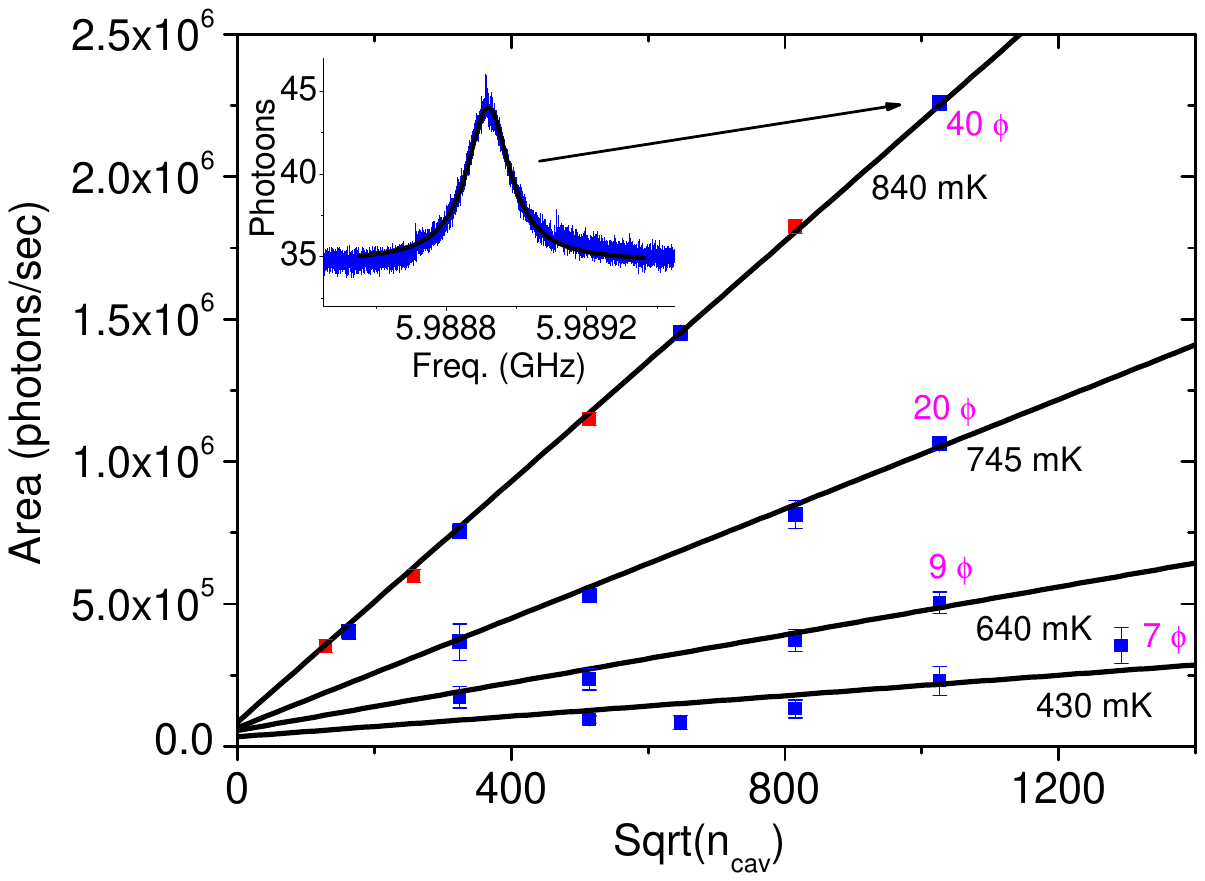}
	\vspace*{-2cm}
			\caption{
			Photon flux due to occupation of the cavity mode in excess of that expected from the cryostat temperature. The $x$ axis is $\sqrt{n_{cav}}$ (blue-detuned pump scheme for blue symbols, red-detuned pump for red). Inset: example of resonance peak obtained from spectrum analyzer. The lines are guides to the eye. The numbers in magenta correspond to the calculated intra-cavity population for the last point of each temperature series (from fit value of $\kappa_{ext}$, see text). }
			\label{fig_cavityheat}
		\end{figure}

The cavity parameters are extracted from a transmission response measurement (amplitude of $S_{21}$ component of $\left[S\right]$ matrix).
This is measured by applying a very weak probe tone (see boxed element in Fig. \ref{fig_wiring}) while keeping the strong pump on. 
We verified that the probe power was small enough not to alter the measurement.
The data (see Fig. \ref{fig_1}) is fit to the expression \cite{RefS12fitAlessandro}:
\begin{equation}
\left( S_{12}\left[\omega\right] \right)_{dB} = 20 \log \left| 1- \frac{Q_{tot}  \left( Q_{ext}+ I Q_{im} \right)^{-1}}{1+2 I Q_{tot} \frac{\omega - \omega_c}{\omega_c}} \right| , \label{eq_S21}
\end{equation}
with $I$ the pure imaginary, $Q_{tot}=\omega_c/\kappa_{tot}$ and $Q_{ext}=\omega_c/\kappa_{ext}$. $Q_{im}$ corresponds to an imaginary component of the external impedance, usually attributed to the inductive bonding wires (leading to the asymmetry of the fit in Fig. \ref{fig_1} center).

		\begin{figure}[h!]
		\centering
	\includegraphics[width=11.5cm]{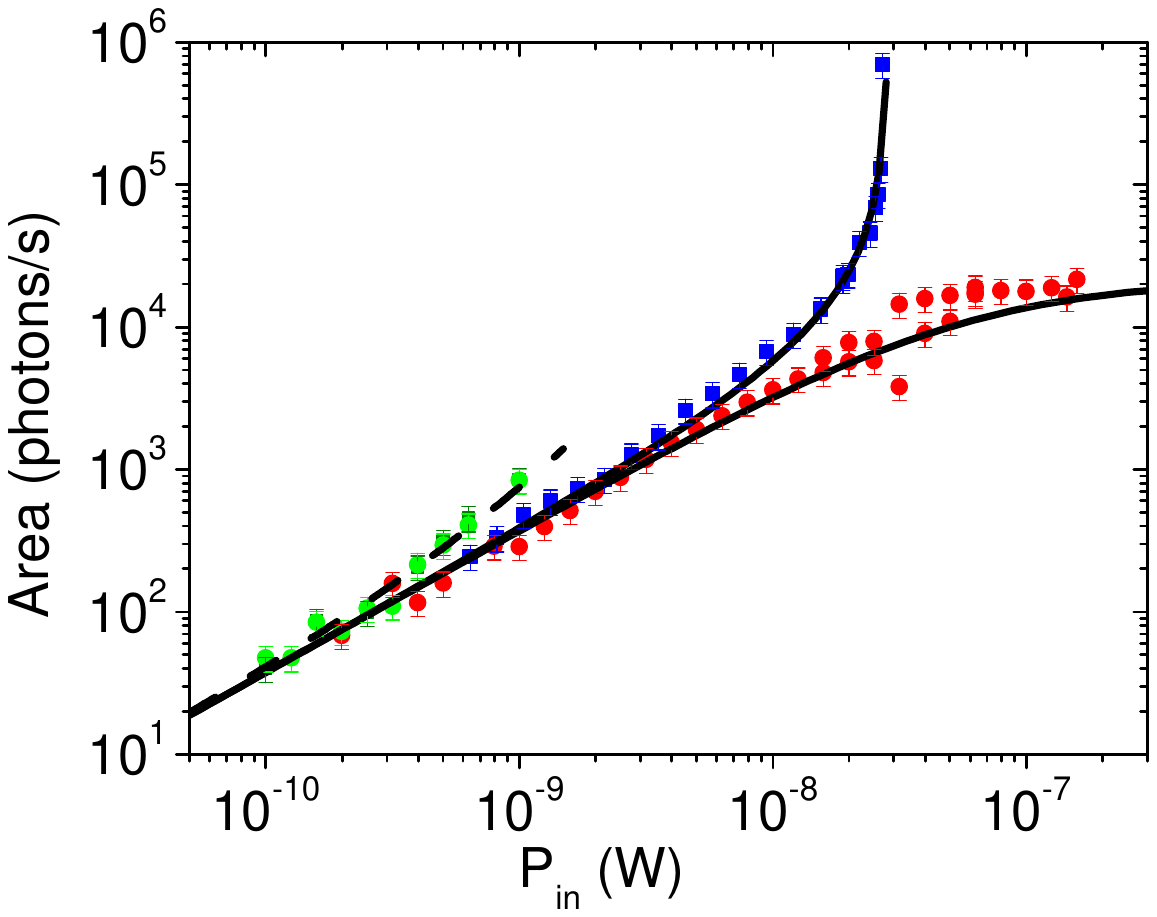}
	\vspace*{-2cm}
			\caption{
			Measured signal amplitude for the 3 schemes used at 210$~$mK: blue-detuned pump, red-detuned pump and ``in-cavity'' (green, the square and circle symbols stand for Stokes and Anti-Stokes peaks respectively). For blue/red pumping, the fits correspond to Eqs. (\ref{eq2}-\ref{eqgain}), defining the coefficient $\cal M$. The dashed line corresponds to the heating measured in Fig. \ref{fig_6}. }
			\label{fig_fit}
		\end{figure}

The fits are always of good quality.
$\kappa_{ext}$ is very stable and reproducible, and does not vary by more than about $\pm 5~\%$.
On the other hand $\kappa_{tot}=\kappa_{ext}+\kappa_{in}$ varies between typically $120~$kHz and $190~$kHz from one cool-down to the other. 
$\kappa_{in}$ represents the internal decay processes, taking into account all microscopic mechanisms.
At low pump powers, $\kappa_{tot}$ happens to be worse (i.e. larger) than at high powers. 
This is true up to some maximum above which we start to substantially degrade the cavity's properties; in terms of $n_{cav}$ this limit lies in the range of $10^8 - 10^9$ photons.
In the range studied the temperature dependence of $\kappa_{tot}$ is rather small. It is immune to the magnetic fields and voltages applied within the parameter range studied.
However, the {\it resonance position} of the cavity is very sensitive to all parameters: so the main issue of the fit is to define $\omega_c$ as accurately as possible.

The properties of the cavity are too complex to allow an analysis of the type of Fig. \ref{fig_6} (performed for the NEMS), which could tell us by how much the microwave mode (and/or the chip) heats at a given pump power. 
We therefore measured directly the thermal population of the cavity with respect to $P_{in}$.
We present these data in Fig. \ref{fig_cavityheat} as a function of $\sqrt{n_{cav}}$: on pure phenomenological grounds, the dependence seems then to be linear (see guides for the eyes in Fig. \ref{fig_cavityheat}). We could not perform this measurement below typically 400$~$mK because the signal was too weak.
The power-dependence is rather different from Fig. \ref{fig_6} and does not seem to be a true heating of the chip itself;
it has thus to be noise fed into the cavity by the pump generator. Indeed, similar features but with different levels of cavity populations have been measured using different brands of generators.
The relevant outcome of this graph is the {\it extra (out-of-equilibrium) cavity population} induced by the strong pump; this is the number quoted in magenta in Fig. \ref{fig_cavityheat} for each temperature, at the largest powers displayed. This number is obtained by dividing the photon flux by $\kappa_{ext}/2$ (bi-directional coupling). Injecting it in the theoretical expressions \cite{AKMreview}, we find out that this effect remains always negligible with respect to other problems creating the finite error bars of Fig. \ref{fig_7}. 

From the careful measurement of the cavity at each point, the knowledge of the heating effects in Figs. \ref{fig_6} and \ref{fig_cavityheat} we can guarantee that the experiment is performed in the best possible conditions. 
To be then quantitative, we finally need to be able to convert photon population in the Stokes/Anti-Stokes peaks into phonon populations.
To do so, we need to know the parameter $\cal M$ introduced in Eq. (\ref{eq2}).
In principle, one could calculate it from theory, knowing all experimental details of the setup \cite{AKMreview}.
However, since all of these parameters are known within some experimental error, the final value obtained for $\cal M$ would be 
of poor precision. It is thus much more efficient to {\it calibrate it} at a given temperature: this is performed in Fig. \ref{fig_fit}, at sufficiently high $T_{cryo}$ to guarantee a good thermal coupling and enough resolution in signal (210$~$mK).
In this plot we show the signal (area) obtained for the 3 schemes presented in Fig. \ref{fig_2}, as a function of $P_{in}$ (they thus all overlap at low powers). The lines are fit to theory Eqs. (\ref{eq1},\ref{eqgain}) \cite{AKMreview}, with the dashed one taking into account the heating produced by the large $n_{cav}$ reached with the ``in-cavity'' scheme. We thus obtain ${\cal M }\approx 1.8 \times 10^{12}~$photons/s/W/K for our device.
Note that at the highest powers used, in the red-detuned pumping scheme the mode cooled from $210~$mK down to approximately $20~$mK.

		\begin{figure}[h!]
		\centering
	\includegraphics[width=11cm]{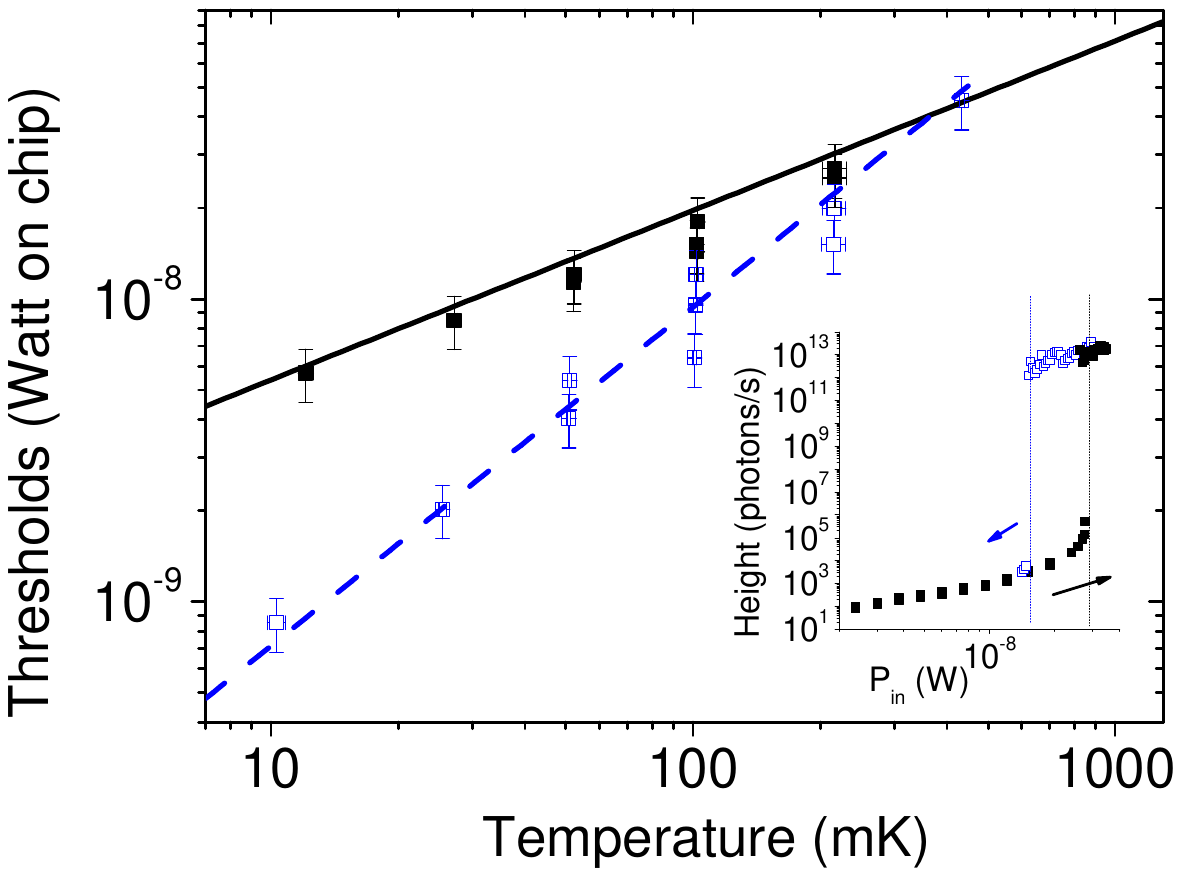}
	\vspace*{-2cm}
			\caption{
			Main: Power threshold $P_{thr}$ to self-oscillation regime (blue-detuned pump scheme). A large hysteresis is seen, certainly due to nonlinearities in the system. The black line is the calculated value $\propto \gamma_m$ from Eq. (\ref{eq1}), and the dashed line is a guide to the eyes $\propto \gamma_m^2$. Inset: peak height measured sweeping the power up (full black), and down (empty blue symbols) at 210$~$mK. The dashed verticals are the threshold positions. }
			\label{fig_oscill} 
		\end{figure}

Our aim was to demonstrate the capabilities of the method for temperatures as low as possible and a coupling $g_0$ which is {\it not} especially large. 
We therefore relied on the out-of-plane flexure of our NEMS beam; in-plane and out-of-plane modes were both characterized first at room temperature using optics \cite{ArcizetThanks}. 
The out-of-plane flexural mode was found at 3.660$~$MHz, shifted about 280$~$kHz from the in-plane one. Both had a quality factor of about $6\,500$.
At low temperatures, the method then relies on the blue-detuned pump instability: below the threshold it enables {\it amplification} of the weak signal, and the position of the threshold itself gives access to the parameter $\gamma_m$ needed for quantitative fits. 
It is thus in principle {\it not necessary} to be able to resolve the signal at low powers; in particular, in these conditions sideband-assymetry thermometry would be impractical.

The definition of $\gamma_m$ from the threshold position is illustrated in Fig. \ref{fig_oscill}.
Ramping up the power, from Eq. (\ref{eq1}) we can recalculate the damping rate from the position of the threshold $P_{thr}$ (full symbols and full line).
Furthermore, it happens that this threshold towards self-sustained oscillations is hysteretic (Fig. \ref{fig_oscill} empty symbols). 
This effect is not yet modeled, but it clearly arises from non-linearities in the system. For microwave optomechanics, this are genuine {\it geometrical} nonlinearites, as opposed to thermal nonlinearities seen in optics \cite{faveroOscill}.
The temperature-dependence of the down-sweep threshold (large amplitude motion) is even stronger than for the up-sweep one (small amplitude).
Again, no thermal decoupling can be seen down to 10$~$mK; the heating effect (Fig. \ref{fig_6}) has been taken into account, and has only a small impact on the data.

\newpage
\begin{widetext}
\vspace*{2cm}

\begin{center} 
{\Large \bf Erratum: On-chip thermometry for microwave optomechanics implemented in a nuclear demagnetization cryostat \\
Phys. Rev. Applied {\bf 12}, 044066 – Published 29 October 2019 \\}
\end{center}
\vspace*{0.5cm}

\begin{center} 
{X. Zhou$^{*,**}$, D. Cattiaux$^{*}$, R. R. Gazizulin$^{*}$, A. Luck$^{*}$, O. Maillet$^{*}$, T. Crozes$^{*}$, J-F. Motte$^{*}$, O. Bourgeois$^{*}$, A. Fefferman$^{*}$ and E. Collin$^{*,\dag}$}
\end{center}

\begin{center}
{\it  (*) Univ. Grenoble Alpes, Institut N\'eel - CNRS UPR2940, 
25 rue des Martyrs, BP 166, 38042 Grenoble Cedex 9, France \\
          (**) Since 01/10/2017: IEMN, Univ. Lille - CNRS UMR8520, 
Av. Henri Poincar\'e, Villeneuve d'Ascq 59650, France}
\end{center}
\vspace*{1cm}

At the time of the work reported in this Article, two of our opto-mechanical samples have been packaged in identical cells; these samples were very similar, apart from the {\it nature of the microwave cavities}. In one cell, these microwave cavities were meanders (as shown in Fig. 1 center) while in the other they were $\lambda$/4 coplanar resonators (similar to Ref. [5]). In both cases, we had 3 cavities coupled evanescently to a transmission line [visible in Fig. 1 (a)], in two of which we had a 50$~\mu$m long nano-mechancial beam. The resonance frequencies of the cavities had been designed to be almost identical for the two setups (about 6.00$~$GHz, 6.25$~$GHz and 6.45$~$GHz). It turns out that our cells have been swapped, such that the cavity picture in Fig. 1 (b) is {\it incorrect}: it should be the coplanar one.  As a consequence, the cavity characteristic impedance is $50~\Omega$ (instead of about 500$~\Omega$ for the meander).
The reported coupling parameter $g_0$ is correct (about $g_0 \approx 0.55~$Hz), which therefore means that the 
mechanical mode that is dealt with is the {\it in-plane one}.
Indeed the $g_0$ of the out-of-plane first flexure is expected about 10 times weaker, which makes it immeasurable. These are the only aspects that have to be modified from the original manuscript. None of the claims of the paper are affected by this fact, and all the quoted numbers and results are accurate.
\vspace*{8cm}

\end{widetext}


\vspace*{2cm}


\begin{thebibliography}{blaaaaaaaaaaaaaaaaaaaaaaaaaaaaaaaaaaaaaaaaaaaa}

\bibitem{EnssPickettEMP} G. Pickett, C. Enss, The European Microkelvin Platform, Nature Review Materials 3, 18012 (2018).
\bibitem{roukescleland} A. N. Cleland and M. L. Roukes, Fabrication of high frequency nanometer scale mechanical resonators from bulk Si crystals, Appl. Phys. Lett. 69, 2653 (1996). 
\bibitem{cleland2010} A. D. O'Connell, M. Hofheinz, M. Ansmann, Radoslaw C. Bialczak, M. Lenander, Erik Lucero, M. Neeley, D. Sank, H. Wang, M. Weides, J. Wenner, John M. Martinis and A. N. Cleland, Quantum ground state and single-phonon control of a mechanical resonator, Nature 464, 697-703 (2010).
\bibitem{quantelecsimmonds} T. A. Palomaki, J. W. Harlow, J. D. Teufel, R. W. Simmonds, K. W. Lehnert, Coherent state transfer between itinerant microwave fields and a mechanical oscillator, Nature 495, 210 (2013).
\bibitem{quantelec2} J.-M. Pirkkalainen, S. U. Cho, Jian Li, G. S. Paraoanu, P. J. Hakonen and M. A. Sillanp\"a\"a, Hybrid circuit cavity quantum electrodynamics with a micromechanical resonator, Nature 494, 211 (2013).
\bibitem{lehnert2008} C. A. Regal, J. D. Teufel And K. W. Lehnert, Measuring nanomechanical motion with a microwave cavity interferometer, Nat. Phys. 4, 555 (2008).
\bibitem{convOskar} Alfredo Rueda, Florian Sedlmeir, Michele C. Collodo, Ulrich Vogl, Birgit Stiller, Gerhard Schunk, Dmitry V. Strekalov, Christoph Marquardt, Johannes M. Fink, Oskar Painter, Gerd Leuchs, and Harald G. L. Schwefel, Efficient microwave to optical photon conversion: an electro-optical realization, Optica 3, 597 (2016).
\bibitem{cindy} A. P. Higginbotham, P. S. Burns, M. D. Urmey, R. W. Peterson, N. S. Kampel, B. M. Brubaker, G. Smith, K. W. Lehnert and C. A. Regal , Harnessing electro-optic correlations in an efficient mechanical converter, Nature Physics 14, 1038 (2018). 
\bibitem{clerknonrec} A. Metelmann and A. A. Clerk, Nonreciprocal Photon Transmission and Amplification via Reservoir Engineering, Phys. Rev. X 5, 021025 (2015).
\bibitem{simmondsamplifnonrecip} Gabriel A. Peterson, Florent Q. Lecocq, Katarina Cicak, Raymond W. Simmonds, Jose A. Aumentado, John D. Teufel, Demonstration of efficient nonreciprocity in a microwave optomechanical circuit, Phys. Rev. X 7, 031001 (2017).
\bibitem{kippenbergamplifnonrecip} N. R. Bernier, L. D. T\'oth, A. Koottandavida, M. A. Ioannou, D. Malz, A. Nunnenkamp, A. K. Feofanov and T. J. Kippenberg, Nonreciprocal reconfigurable microwave optomechanical circuit, Nature Comm. 8, 604 (2017).
\bibitem{jfink} S. Barzanjeh, M. Wulf, M. Peruzzo, M. Kalaee, P.B. Dieterle, O. Painter and J.M. Fink, Mechanical on-chip microwave circulator, Nature Comm. 8, 953 (2017).
\bibitem{schwavsciencesqueeze} E. E. Wollman, C. U. Lei, A. J. Weinstein, J. Suh, A. Kronwald, F. Marquardt, A. A. Clerk, K. C. Schwab, Quantum squeezing of motion in a mechanical resonator, Science 349, 952 (2015).
\bibitem{sillanpaaintrique} C. F. Ockeloen-Korppi, E. Damsk\"agg, J.-M. Pirkkalainen, M. Asjad, A. A. Clerk, F. Massel, M. J. Woolley and M. A. Sillanp\"a\"a, Stabilized entanglement of massive mechanical oscillators, Nature 556, 478 (2018).
\bibitem{sillanpaamultimode} Francesco Massel, Sung Un Cho, Juha-Matti Pirkkalainen, Pertti J. Hakonen, Tero T. Heikkil\"a and Mika A. Sillanp\"a\"a, Multimode circuit optomechanics near the quantum limit, Nat. Comm. 3, 987 (2012).
\bibitem{schwabtones} A. J. Weinstein, C. U. Lei, E. E. Wollman, J. Suh, A. Metelmann, A. A. Clerk, and K. C. Schwab, Observation and Interpretation of Motional Sideband Asymmetry in a Quantum Electromechanical Device, Phys. Rev. X 4, 041003 (2014).
\bibitem{ArmourBlencowe} A. D. Armour and M. P. Blencowe, Probing the quantum coherence of a nanomechanical resonator using a superconducting
qubit: I. Echo scheme, New J. of Phys. 10, 095004 (2008). 
\bibitem{bouwmeesterNJP} Dustin Kleckner, Igor Pikovski, Evan Jeffrey, Luuk Ament, Eric Eliel, Jeroen van den Brink, and Dirk Bouwmeester, Creating and verifying a quantum superposition in a micro-optomechanical system, New J. of Phys. 10, 095020 (2008).
\bibitem{KlecknerThesis} Dustin Paul Kleckner, {\it Micro-Optomechanical Systems for Quantum Optics}, PhD Thesis of University of California, Santa Barbara (04/2010).
\bibitem{AKMreview} Markus Aspelmeyer, Tobias J. Kippenberg, Florian Marquardt, Cavity optomechanics, Rev. of Mod. Phys. 86, 1391 (2014).
\bibitem{noiseJohn} A. Shibahara, O. Hahtela, J.Engert, H. van der Vliet, L. V. Levitin, A.Casey, C.P.Lusher, J. Saunders, D. Drung, and Th. Schurig, Primary current-sensing noise thermometry in the millikelvin regime, Trans. R. Soc. A 374, 20150054 (2016).
\bibitem{Rob3Hefork} R. Blaauwgeers, M. Blazkova, M. Clovecko, V.B. Eltsov, R. de Graaf, J. Hosio, M. Krusius, D. Schmoranzer, W. Schoepe, L. Skrbek, P. Skyba, R.E. Solntsev, D.E. Zmeev, Quartz Tuning Fork: Thermometer, Pressure- and Viscometer for Helium Liquids, Journal of Low Temperature Physics 146, 537 (2007).
\bibitem{roukesmass} A. K. Naik, M. S. Hanay, W. K. Hiebert, X. L. Feng and M. L. Roukes, Towards single-molecule nanomechanical mass spectrometry, Nat. Nanotech. 4, 445 (2009).
\bibitem{sillanpaPrivCom} M. A. Sillanp\"a\"a, private communication.
\bibitem{schwabPHD} Tristan Rocheleau, {\it Quantum-Limited Mechanical Resonator Measurement and Back-Action Cooling to near the Quantum Ground State}, PhD thesis of Cornell University (2011).
\bibitem{contactLehnert} K. Lehnert and J. Teufel, private communication.
\bibitem{jfinkpriv} S. Barzanjeh, private communication. 
\bibitem{TeufelBEAM} J. D. Teufel, T. Donner, M. A. Castellanos-Beltran, J. W. Harlow, and K.W.Lehnert, Nanomechanical motion measured with an imprecision below that at the standard quantum limit, Nature Nanotech. 4, 820 (2009).
\bibitem{schwabrocheleau} T. Rocheleau, T. Ndukum, C. Macklin, J. B. Hertzberg, A. A. Clerk and K. C. Schwab, Preparation and detection of a mechanical resonator near the ground state of motion, Nature 463, 72 (2010).
\bibitem{DavisTLS} B. D. Hauer, P. H. Kim, C. Doolin, F. Souris, and J. P. Davis, Two-level system damping in a quasi-one-dimensional optomechanical resonator, Phys. Rev. B 98, 214303 (2018).
\bibitem{KunalsJLTP} M. Defoort, K.J. Lulla, C. Blanc, H. Ftouni, O. Bourgeois, and E. Collin, Stressed Silicon Nitride Nanomechanical Resonators at Helium Temperatures, J. of Low Temp. Phys. 171, 731 (2013).
\bibitem{paperYuriSatge} C. B\"auerle, Y. Bunkov, S.N. Fisher, Chr. Gianese and H. Godfrin, The new Grenoble 100 microKelvin refrigerator,
Czekoslovak J. of Phys. {\bf 46}, suppl S5, 2791-2792, (1996).
\bibitem{firstopticsSBcool} Jasper Chan, T. P. Mayer Alegre, Amir H. Safavi-Naeini, Jeff T. Hill, Alex Krause, Simon Gr\"oblacher, Markus Aspelmeyer and Oskar Painter, Laser cooling of a nanomechanical oscillator into its quantum ground state, Nature 478, 89 (2011).
\bibitem{microwaveSBCoolLehnertPRL} J. D. Teufel, Tobias Donner, Dale Li, J. W. Harlow, M. S. Allman, Katarina Cicak, A. J. Sirois, Jed D. Whittaker, K. W. Lehnert, Raymond W. Simmonds, Sideband cooling of micromechanical motion to the quantum ground state, Nature 475, 359 (2011).
\bibitem{clerkmarquardtharris2010} Florian Marquardt, J. G. E. Harris, and S. M. Girvin, Dynamical Multistability Induced by Radiation Pressure
in High-Finesse Micromechanical Optical Cavities, Phys. Rev. Lett. 96, 103901 (2006).
\bibitem{kippenbergOscillVahalaPRL2005} T. J. Kippenberg, H. Rokhsari, T. Carmon, A. Scherer, and K. J. Vahala, Analysis of Radiation-Pressure Induced Mechanical Oscillation of an Optical Microcavity, Phys. Rev. Lett. 95, 033901 (2005).
\bibitem{delftsteeleOscill} V. Singh, S. J. Bosman, B. H. Schneider, Y. M. Blanter, A. Castellanos-Gomez and G. A. Steele, Optomechanical coupling between a multilayer graphene mechanical resonator and a superconducting microwave cavity, Nat. Nanotech. 9, 820 (2014).
\bibitem{TLSMohanty} P. Mohanty, D. A. Harrington, K. L. Ekinci, Y. T. Yang, M. J. Murphy, and M. L. Roukes, Intrinsic dissipation in high-frequency micromechanical resonators, Phys. Rev. B 66, 085416 (2002).
\bibitem{TLSPashkin} F. Hoehne, Yu. A. Pashkin, O. Astafiev, L. Faoro, L. B. Ioffe, Y. Nakamura, and J. S. Tsai, Damping in high-frequency metallic nanomechanical resonators, Phys. Rev. B 81, 184112 (2010).
\bibitem{KunalPRB} A. Venkatesan, K. J. Lulla, M. J. Patton, A. D. Armour, C. J. Mellor, and J. R. Owers-Bradley, Dissipation due to tunneling two-level systems in gold nanomechanical resonators, Phys. Rev. B 81, 073410 (2010).
\bibitem{KunalPRL} K. J. Lulla, M. Defoort, C. Blanc, O. Bourgeois, and E. Collin, Evidence for the role of normal-state electrons in nanoelectromechanical damping mechanisms at very low temperatures, Phys. Rev. Lett. 110, 177206 (2013).
\bibitem{FaveroTLSPRL} M. Hamoumi, P. E. Allain, W. Hease, E. Gil-Santos, L. Morgenroth, B. G\'erard, A. Lema\^itre, G. Leo, and I. Favero, Microscopic Nanomechanical Dissipation in Gallium Arsenide Resonators, Phys. Rev. Lett. 120, 223601 (2018). 
\bibitem{Painter}  Gregory S. MacCabe, Hengjiang Ren, Jie Luo, Justin D. Cohen, Hengyun Zhou, Alp Sipahigil, Mohammad Mirhosseini, and Oskar Painter, Phononic bandgap nano-acoustic cavity with ultralong phonon lifetime, arXiv:1901.04129v1 (2019).
\bibitem{RoukesLifshitzClamp} M. C. Cross and Ron Lifshitz, Elastic wave transmission at an abrupt junction in a thin plate with application to heat transport and vibrations in mesoscopic systems, Phys. Rev. B 64, 085324 (2001).
\bibitem{schwab7mK} J. Suh, A. J. Weinstein, C. U. Lei, E. E. Wollman, S. K. Steinke, P. Meystre, A. A. Clerk, K. C. Schwab, Mechanically Detecting and Avoiding the Quantum Fluctuations of a Microwave Field, Science 344, 1262 (2014).
\bibitem{vortexNoise} C. Song, T.W. Heitmann, M.P. DeFeo, K. Yu, R. McDermott, M. Neeley, John M. Martinis, and B.L.T. Plourde, Microwave response of vortices in superconducting thin films of Re and Al, Phys. Rev. B 79, 174512 (2009).
\bibitem{chargeNoise} A. B. Zorin, F.-J. Ahlers, J. Niemeyer, T. Weimann, H. Wolf, V. A. Krupenin, and S. V. Lotkhov, Background charge noise in metallic single-electron tunneling devices, Phys. Rev. B 53, 13682 (1996).
\bibitem{adsorbate} S.E. de Graaf, L. Faoro, J. Burnett, A.A. Adamyan, A.Ya. Tzalenchuk, S.E. Kubatkin, T. Lindstr\"om and A.V. Danilov, Suppression of low-frequency charge noise in superconducting resonators by surface spin desorption, Nature Comm. 9, 1143 (2018).
\bibitem{dielecTLS} Jiansong Gao, Miguel Daal, John M. Martinis, Anastasios Vayonakis, Jonas Zmuidzinas, Bernard Sadoulet, Benjamin A. Mazin, Peter K. Day, and
Henry G. Leduc, A semiempirical model for two-level system noise in superconducting microresonators, Appl. Phys. Lett. 92, 212504 (2008).
\bibitem{fongPRB} King Y. Fong, Wolfram H. P. Pernice, and Hong X. Tang, Frequency and phase noise of ultrahigh Q silicon nitride nanomechanical resonators, Phys. Rev. B 85, 161410(R) (2012).
\bibitem{OlivierPhD} Olivier Maillet, {\it Stochastic and non-linear processes in nano-electro-mechanical systems}, PhD Thesis of Universit\'e Grenoble Alpes (25/05/2016).
\bibitem{ACSnanoUs} Olivier Maillet, Xin Zhou, Rasul R. Gazizulin, Bojan R. Ilic, Jeevak M. Parpia, Olivier Bourgeois, Andrew D. Fefferman, and Eddy Collin,
Measuring frequency fluctuations in nonlinear nanomechanical resonators, ACS Nano 12, 5753 (2018). 
\bibitem{RoukesHentzNatNanotech} Marc Sansa, Eric Sage, Elizabeth C. Bullard, Marc G\'ely, Thomas Alava, Eric Colinet, Akshay K. Naik, Luis Guillermo Villanueva, Laurent Duraffourg, Michael L. Roukes, Guillaume Jourdan, S\'ebastien Hentz, Frequency fluctuations in silicon nanoresonators, Nature Nanotech. 11, 552-558 (2016)
\bibitem{ustinov} Alexander  Bilmes, Sebastian  Zanker, Andreas  Heimes, Michael  Marthaler, Gerd  Sch\"on, Georg  Weiss, Alexey  V.  Ustinov, and  J\"urgen  Lisenfeld, Electronic Decoherence of Two-Level Systems in a Josephson Junction, Physical review B 96, 064504 (2017).
\bibitem{OliveNJP} O. Maillet, F. Vavrek, A.D. Fefferman, O. Bourgeois and E. Collin, Classical decoherence in a nanomechanical resonator, New J. Phys. 18, 073022 (2016).
\bibitem{MartialPhD} M. Defoort, {\it Non-linear dynamics in nano-electromechanical systems at low temperatures}, PhD Thesis of Universit\'e Grenoble Alpes (16/12/2014).
\bibitem{ParpiaGamma} E. Nazaretski, R. D. Merithew,V.O. Kostroun, A.T. Zehnder, R.O. Pohl, and J.M. Parpia, Effect of Low-Level Radiation on the Low Temperature Acoustic Behavior of {\it a}-SiO$_2$, Phys. Rev. Lett. 92, 245502-1 (2004).
\bibitem{ULANCHSpaper} N. S. Lawson, A simple heat switch for use at millikelvin temperatures, Cryogenics  22, 667 (1982).
\bibitem{RefHenriULTS} R. Rusby, D. J. Cousins, D. Head, P. Mohandas, Yu. M. Bunkov, C. Bäuerle, R. Harakaly,
E. Collin, S. Triqueneaux, C. Lusher, J. Li, J. Saunders, B. Cowan, J. Nyéki, M. Digby,
J. Pekola, K. Gloos, P. Hernandez, M. de Groot, A. Peruzzi, R. Jochemsen, A. Chinchure,
W. Bosch, F. Mathu, J. Flokstra, D. Veldhuis, Y. Hermier, L. Pitre, A. Vergé,
F. Benhalima, B. Fellmuth, J. Engert, {\it Dissemination of the Ultra-Low Temperature Scale PLTS-2000}, Proceedings of TEMPMEKO, Vols. 1 \& 2, B. Fellmuth, J. Seidel, G. Scholz (ed.), Berlin, VDE Verlag GmbH (2002)
\bibitem{PobellBook} F. Pobell, {\it Matter and methods at low temperatures}, Springer-Verlag Berlin Heidelberg, Second Edition (1996).
\bibitem{GuenaultViW} A.M. Gu\'enault, V. Keith, C.J. Kennedy, S.G. Mussett, and G.R. Pickett. The mechanical behavior of a vibrating
wire in superfluid $^3$He-B in the ballistic limit, J. of Low Temp. Phys. 62, 511 (1986).
\bibitem{ClemensViW} C. B. Winkelmann, E. Collin, Yu. M. Bunkov, H. Godfrin, Vibrating wire thermometry in superfluid $^3$He, J. of Low Temp. Phys.135, 3  (2004).
\bibitem{ULANCFork}  D.I. Bradley, M. Clovecko, S.N. Fisher, D. Garg, A. Guénault, E. Guise, R.P. Haley, G.R. Pickett, M. Poole, V. Tsepelin, Thermometry in Normal Liquid He-3 Using a Quartz Tuning Fork Viscometer, J. of Low Temp. Phys. 171, 750 (2013).
\bibitem{ULANCFork2} D.I. Bradley, P. Crookston, S.N. Fisher, A. Ganshin, A. Guénault, R.P. Haley, M.J. Jackson, G.R. Pickett, R. Schanen, V. Tsepelin, The damping of a quartz tuning fork in superfluid He-3-B at low temperatures, J. of Low Temp. Phys. 157, 476 (2009).
\bibitem{Book3HeVollhardt}  Dieter Vollhardt and Peter Wolfle, {\it The Superfluid Phases of Helium 3},  Dover Books on Physics, NY (2013). 
\bibitem{HookHall1} D.C. Carless, H.E. Hall, and J.R. Hook, Vibrating wire measurements in liquid $^3$He. I. The normal state, J. of Low Temp. Phys. 50, 583 (1983). 
\bibitem{HookHall2} D.C. Carless, H.E. Hall, and J.R. Hook, Vibrating wire measurements in liquid $^3$He. II. The superfluid B phase, J. of Low Temp. Phys. 50, 605 (1983).
\bibitem{RefS12fitAlessandro} M. S. Khalil, M. J. A. Stoutimore, F. C. Wellstood, and K. D. Osborn, An analysis method for asymmetric resonator transmission applied to superconducting devices, J. Appl. Phys. 111, 054510 (2012).
\bibitem{ArcizetThanks} A room-temperature vacuum chamber equipped with a laser readout from O. Arcizet and B. Pigeau has been used.
\bibitem{faveroOscill} Christophe Baker, Sebastian Stapfner, David Parrain, Sara Ducci, Giuseppe Leo,
Eva M. Weig and Ivan Favero, Optical Instability and Self-Pulsing in Silicon Nitride Whispering Gallery Resonators, Optics Express 20, 29076 (2012).

\end{thebibliography}
\end{document}